\documentclass[preprint]{aastex}
\def\simgr{\,\hbox{\hbox{$ > $}\kern -0.8em \lower 1.0ex\hbox{$\sim$}}\,}
\def\simle{\,\hbox{\hbox{$ < $}\kern -0.8em \lower 1.0ex\hbox{$\sim$}}\,}

\shortauthors{THORSTENSEN et al.}
\shorttitle{SDSS Cataclysmics}

\begin{document}
\title{Spectroscopic Orbital Periods for 29 Cataclysmic 
Variables from the Sloan Digital Sky Survey
}

\author{John R. Thorstensen, Cynthia J. Taylor, Christopher S. Peters, 
and Julie N. Skinner}
\affil{Department of Physics and Astronomy\\
6127 Wilder Laboratory, Dartmouth College\\
Hanover, NH 03755-3528}

\author{John Southworth}
\affil{Astrophysics Group\\ 
Keele University\\
Staffordshire ST5 5BG, UK}

\author{Boris T. G\"ansicke}
\affil{Department of Physics\\ 
University of Warwick\\ 
Coventry CV4 7AL, UK}

\begin{abstract}
We report follow-up spectroscopy of 29 cataclysmic variables
from the Sloan Digital Sky Survey (SDSS), 22 of which were discovered
by SDSS and seven other previously known systems that were recovered in SDSS.
The periods for 16 of these objects were included in the tabulation
by \citet{unveils}.  While most of the systems have periods less
than 2 hours, only one has a period in the 80-86 minute `spike'
found by \citet{unveils}, and 11 have periods longer than 
3 hours, indicating that the present sample is skewed toward longer-period,
higher-luminosity objects.  Seven of the objects have spectra resembling 
dwarf novae, but have apparently never been observed in
outburst, suggesting that many cataclysmics
with relatively low variability amplitude remain to be 
discovered.  Some of the objects are notable.
SDSS J07568+0858 and SDSS J08129+1911 were previously known to have
deep eclipses; in addition to spectroscopy, we use archival
data from the CRTTS to refine their periods.  
We give a parallax-based distance of 195 (+54, $-$39) pc 
for LV Cnc (SDSS J09197+0857), which at $P_{\rm orb} = 81$ m
has the shortest orbital period in our sample.  
SDSS J08091+3814 shows both the spectroscopic
phase offset and phase-dependent absorption found in 
SW Sextantis stars.
The average spectra of SDSS J08055+0720 and SDSS J16191+1351 show 
contributions from K-type secondaries, and SDSS J080440+0239 shows
a contribution from an early M star.  We use these to 
constrain the distances.  
SDSS J09459+2922 has characteristics
typical of a magnetic system.  
SDSS11324+6249 may be a novalike variable, and if so, its
orbital period (99 min) is unusually short for that subclass.

\end{abstract}


\section{Introduction}

Cataclysmic variable stars (CVs) are close binary systems in which
a white dwarf primary star accretes matter by way of Roche
lobe overflow from a companion (the secondary), which usually resembles
a late-type main-sequence star\footnote{Some objects have
white dwarf secondaries (as well as primaries); these are
known as AM CVn stars, and while they are usually considered to be
CVs, we are not including them in this discussion.} \citep{warner95}. 
Because the red-giant precursor
of a white dwarf can become much larger than the 
semimajor axis of a typical CV, it is thought that CVs 
are remnants of common-envelope evolution.  Subsequent evolution
proceeds through loss of orbital angular momentum, probably
through magnetic braking of the co-rotating secondary, and
at the shortest periods through gravitational radiation losses.
For most systems, the loss of angular momentum causes the
orbital period to decrease, but eventually the secondary
becomes degenerate, so that subsequent mass loss causes its 
radius to increase.  If this point has been passed, 
the orbital period should increase gradually
as mass transfer continues.  This evolutionary turnaround,
known as the `period bounce', is predicted to occur at orbital
period $P_{\rm orb}$ near
65 min, though empirically the minimum period appears
to be somewhat longer \citep[$\sim 75$ min; ][]{barker03}. 
Standard evolutionary theory, then, predicts 
that few CVs will be found below the period minimum.  Also,
the gravitational radiation thought to dominate the angular
momentum losses at short periods works rather slowly, so a
relatively large number of CVs should have periods just 
longward of the minimum, forming a ``spike'' in the 
distribution. 

Until recently, observational evidence for this scenario
was somewhat limited.  CVs with hydrogen-rich secondaries
do respect the lower period bound; the
few exceptions appear to be either stars that evolved significantly
before beginning mass transfer \citep{eipsc,v485cen} or unusual
objects of other provenance \citep{sdss1507patt,sdss1507litt}.  
But the distribution
of known CV periods did not show the predicted spike 
\citep{pattlate,barker03,unveils}.

The mediocre agreement with theory prompts us to consider
how CVs are selected.  Until 
recently, CVs were detected mostly through photometric 
variability (especially dwarf nova outbursts), X-ray 
emission, and ultraviolet color.  The Sloan 
Digital Sky Survey turned up a large number of new CVs 
\citep[hereafter collectively referred to as SzkodySDSSCVs]{szkodyi,szkodyii,szkodyiii,szkodyiv,szkodyv,
szkodyvi,szkodyvii}.
CV colors are unusual because of their composite
spectra, with the result that SDSS allotted spectroscopic
fibers to many of them.  Although CV spectra are
heterogeneous, they tend to be fairly distinctive, so many
of the CVs could be classified with high confidence.
The criteria for allotting spectroscopic 
fibers to stellar objects
were mostly designed to generate a 
large sample of galaxies and QSOs, not to observe every available CV,
so there is no assurance that the SDSS CV sample is 
truly complete. 
Nonetheless, the combination of the great
depth of SDSS, and the lack of bias toward outbursting 
objects, led to the discovery of a large number of 
low-luminosity objects just above the period minimum.
\citet{unveils} compiled results for 137 SDSS CVs (most 
of them newly discovered),
showed that their orbital period distribution
was significantly different from previously-known samples,
and identified the short-period objects with the long-predicted
period spike.  

Here we present spectroscopy and orbital
periods for 29 CVs;  Table \ref{tab:star_info} lists the objects and
gives the short names we will use for them here.   
Twenty-two of these were originally discovered
in SDSS data.  We also include seven more objects which
had been known previously but which were recovered in the 
SDSS.  \citet{unveils} listed orbital periods 
for 16 of the objects, and we give the details here. 
Observations of the remaining objects 
were not completed until after the \citet{unveils} paper
was prepared.  The periods for most of these appear here
for the first time.

\section{Observations and Reductions}

\subsection{MDM Observations and Reductions}

Most of the observations reported here are from the 2.4 m
Hiltner telescope at MDM Observatory on Kitt Peak, 
Arizona, equipped with the modular
spectrograph, a 600 line mm$^{-1}$ grating, and a 2048$^2$ 
SITe CCD (the `Echelle' detector).  This gave 2.0 \AA\ pixel$^{-1}$ and a spectral
resolution of $\sim 3.5$ \AA\ from 4210 to
7500 \AA, with strong vignetting toward the ends of the spectral
range.  For some of the later observations, we used
a 1024$^2$ SITe CCD (`Templeton'), which had the same
pixel as `Echelle' and scale and similar sensitivity, but 
less spectral coverage (4660 to 6730 \AA).  The projected slit width 
was $\sim 1.1$ arcsec.  We often observed far from the meridian to resolve
ambiguities in the daily cycle count; when doing this, we
rotated the instrument to position the slit near the 
parallactic angle, to minimize atmospheric dispersion effects
\citet{filippenko}\footnote{The most recent version of JRT's
{\it JSkyCalc} observing planning program includes a tool
for computing the optimum position angle for a spectrograph
slit over an observation interval; it is available at
{\tt http://www.dartmouth.edu/$\sim$physics/faculty/skycalc/flyer.html}.}.
During twilight we took spectra of 
bright early-type stars to derive an approximate
correction for the telluric bands, flux standard
stars when conditions warranted, and  
He, Ne, and Xe lamps for wavelength calibration.  For
spectra taken during the night, we used the $\lambda 5577$
night sky line to establish the velocity zero point.
The night-sky line technique lets us monitor the radial
velocities as they come in, making for very efficient 
use of observing time.  Individual exposures were typically
6-8 minutes, to avoid smearing the velocities excessively
at short periods.

Two of the objects, SDSS0812 and SDSS1028 (Leo5), were bright enough for the
McGraw-Hill 1.3 m telescope, also at MDM.  For this
we used the Mark III spectrograph, the SITe 1024$^2$
CCD `Templeton', and a grism giving 2.2 \AA\ pixel$^{-1}$ 
from 4650 to 6980 \AA , with a spectral resolution of 
4.0 \AA .  

Whenever the sky was clear, we observed flux standard stars
in twilight.  We estimate that the absolute fluxes are 
accurate to about 20 percent, because of transparency
variations and uncalibratable losses at the 1.1-arcsecond
slit.  In addition, the modspec spectra sometimes suffer from 
irreproducible, wavelength-dependent variations in the
response, which can reach 20 or 30 per cent amplitude 
in individual
spectra.  For the most part these average out fairly
well when large numbers of spectra are combined.  
The $V$ magnitudes given in Table \ref{tab:star_info} were
synthesized from our flux-calibrated spectra using the 
iraf {\it sbands} task and the $V$ passband tabulated
by \citet{bessell}.

Our average spectra are included in Figs. 1 through 8
(along with the period searches and folded velocity curves discussed
later).
The line strengths, line profiles, continuum shapes,
and flux levels are mostly similar to those in SzkodySDSSCVs.
The $V$ magnitudes inferred from our 
spectral continua (Table \ref{tab:star_info}) also agree 
fairly well with the $g$ magnitudes from SDSS imaging.  We
therefore observed most of the systems
in states similar to when they were discovered;
exceptions are noted in the discussions of 
individual objects (Section \ref{sec:individuals}).

We reduced the MDM data primarily with IRAF routines, but 
for extracting one-dimensional spectra from the two-dimensional 
images, we used our own implementation of the optimal-extraction
algorithm described by \citet{horne}.  

To measure radial velocities of the H$\alpha$ emission
line, we used an algorithm developed by \citet{sy80}, in
which the line profile is convolved with an antisymmetric
function and the zero of the convolution is taken 
to be the line center.  The antisymmetric function
was usually composed of positive and negative gaussians,
separated by an interval chosen to emphasize the steep
parts of the line profile.  Uncertainties in the velocities
were estimated by propagating the errors of the 
counts in each spectral channel, as estimated from 
photon statistics and read noise; line-profile 
variations can introduce scatter significantly larger
than this.  For absorption-line velocities
we used the {\it rvsao} cross-correlation
package \citep{kurtzmink}; the template was the sum of a large number
of late-type IAU velocity standards shifted to zero velocity.
The {\it rvsao} package estimates velocity uncertainties
using the antisymmetric part of the cross-correlation
\citep{tonrydavis79}. 
Table \ref{tab:velocities} lists all the radial velocities.

To search for periods, we used the `residual-gram'
algorithm described by \citet{tpst}.
We fitted the time series with sinusoids at the
best periods (see \citealt{tf85} for details); 
Table \ref{tab:parameters} lists the parameters
of the curves that are plotted in Figs. 1 -- 8.
Because essentially all our observations
are from a single longitude, the times of observation fall in a 
limited window {\it modulo} one sidereal day, so period searches
inevitably show aliases spaced by 1 cycle d$^{-1}$, 
corresponding to alternate choices of daily cycle count.
We minimized the resulting ambiguities by taking observations
over as wide a range of 
hour angles as possible.  When appropriate, we 
used the Monte Carlo test described by \citet{tf85} to 
estimate the reliability of our choice of alias.

\subsection{WHT Observations and Reductions}

A few of our spectra are from the ISIS Double Beam Spectrograph 
at the William Herschel Telescope at La Palma.  The 
instrument configuration, observation procedure, and reductions were 
essentially as described by \citet{southworth07}, except that we implemented 
the \citet{sy80} radial-velocity measurements using a python-language 
program based on PyFits and NumPy.

\subsection{Distances from Secondary Stars}
\label{sec:secondaries}

The spectra of CVs sometimes show a contribution from a late type
star, which can constrain the distance.
To do this, we use an interactive program to scale and subtract
spectra of stars of known spectral type from
the program star's spectrum, and adjust the spectral type and
scale factor to cancel the late-type features.
The subtracted spectra are either M- and K-dwarfs
classified respectively by \citet{boeshaar} or \citet{keenan}, and 
taken with the same instrument as the program star, or
SDSS M-dwarf templates from \citet{bochanskitemplate}. 
For each spectral type giving reasonable cancellation, we
estimate the acceptable range of scale factors, and
hence the apparent $V$ magnitudes of the secondary.
We use the spectral type to find the surface brightness
in the $V$-band, using data tabulated by \citet{beuermann99}.  

The orbital period, together with the assumption that the secondary
must fill its Roche critical lobe, strongly constrains the
secondary's radius $R_2$ :  at a given $P_{\rm orb}$, 
$R_2$ is roughly proportional to $M_2^{1/3}$, where
$M_2$ is the secondary's mass.  
We seldom can measure $M_2$ directly, but using evolutionary
scenarios \citep{bk00} and the known $P_{\rm orb}$, we constrain $M_2$ 
to a broad range of plausible values, which results in a 
relatively narrow range of radii thanks to the weak dependence.
Combining $R_2$ with the
surface brightness and radius gives an absolute $V$ magnitude.
Comparing this with the apparent $V$ of the secondary's
contribution, and including a correction for interstellar 
extinction \citep[guided by][]{schlegel98} 
gives an estimate of the distance.  Note that 
we do not assume that the secondary is a main-sequence star.  

Because spectral features tend to grow stronger toward later
spectral types, there is often a correlation between the 
secondary's spectral type and the fraction of the
light attributed to it.  We therefore compute the distance and
its uncertainty using a Monte Carlo approach, that is, by 
running the calculation many times, every time selecting
each input value randomly from its possible range.
Our quoted distance is the median of the resulting histogram, and
the range that includes 68 percent of the results is our 
quoted ``1-sigma'' uncertainty.
When picking input values, we select a spectral type and
then pick a magnitude based on the allowed range at that spectral
type.  This explicitly accounts for the correlation.

\section{The Individual Stars}
\label{sec:individuals}
%
%
%
%
\subsection{SDSS J07568+0858} 

\citet{szkodyvii} noted that the SDSS spectrum of this object shows strong 
HeII $\lambda$4686 emission, and suggested it may be an intermediate
polar (i.e., a magnetic system with a white dwarf spin period substantially
shorter than $P_{\rm orb}$).  They also found a preliminary period 
near 2 h.  While the present paper was in preparation, \citet{tov2014} 
presented a comprehensive photometric and spectroscopic study
of the object, finding it to be an eclipsing system with 
$P_{\rm orb} = 0.1369745(4)$ d.  

Our findings largely confirm those of \citet{tov2014}.
Our H$\alpha$ velocities give an unambiguous $P_{\rm orb} = 3.292 \pm 0.003$ h,
which is consistent with the much more precise value they find, and
a velocity amplitude identical to theirs.  By coincidence, the
eclipse epoch given in their ephemeris (HJD 2455958.59184) is less
than two weeks from the epoch of our spectroscopic observations, 
so we can compute eclipse-based phases for our velocities without
significant uncertainty.
In Fig.~1 we show the velocities folded on the eclipse ephemeris. 
There is an intriguing indication of a rotational disturbance, i.e., a displacement
toward positive velocity before eclipse, and a negative 
displacement following eclipse, as expected from the eclipse
of a rotating structure.

The Catalina Real Time Transient 
Survey, Data Release 2 \citep[hereafter CRTTS]{drakecrtts} 
\footnote{The Catalina photometric data are available at 
http://nesssi.cacr.caltech.edu/DataRelease/}
gives 308 measurements of this source from 2005 to 2013.  The source
is mostly around 16.3 mag, but with occasional excursions down to fainter
than 18th -- similar to the eclipse depth seen by \citet{tov2014}.  To 
see whether the fainter points were consistent with the eclipse, we 
downloaded the CRTTS data, corrected the timings to the solar
system barycenter, and searched for periodicities near
the known eclipse period using a `string length'
algorithm \citep{dworetsky83}.
The \citet{tov2014} period is strongly detected, along with 
weak aliases corresponding to the annual cycle count.   

The long time span of the CRTTS data allowed us to refine
the period.  The isolated observations from CRTTS do not permit
the extraction of individual eclipse timings.  To constrain
the period, we instead
varied the period slightly and examined the effect on folded
light curves.  As one might expect, the eclipse appeared
sharply defined only for a narrow range of periods, namely 
0.1369739(2) d. 
The central value is only 1.5 standard deviations from the \citet{tov2014}
period, which is 0.1369745(4) d.  Fig.~\ref{fig:crtscurves} shows the 
CRTTS magnitudes folded using the refined period.


\subsection{SDSS J08055+0720} 

The spectrum of this object shows a strong contribution from a secondary star
along with emission lines and a continuum typical of dwarf
novae at minimum light.  However, the CRTTS
shows the source near 18th magnitude without any outbursts detected despite 
good coverage over eight observing seasons.  

\begin{figure}
\plotone{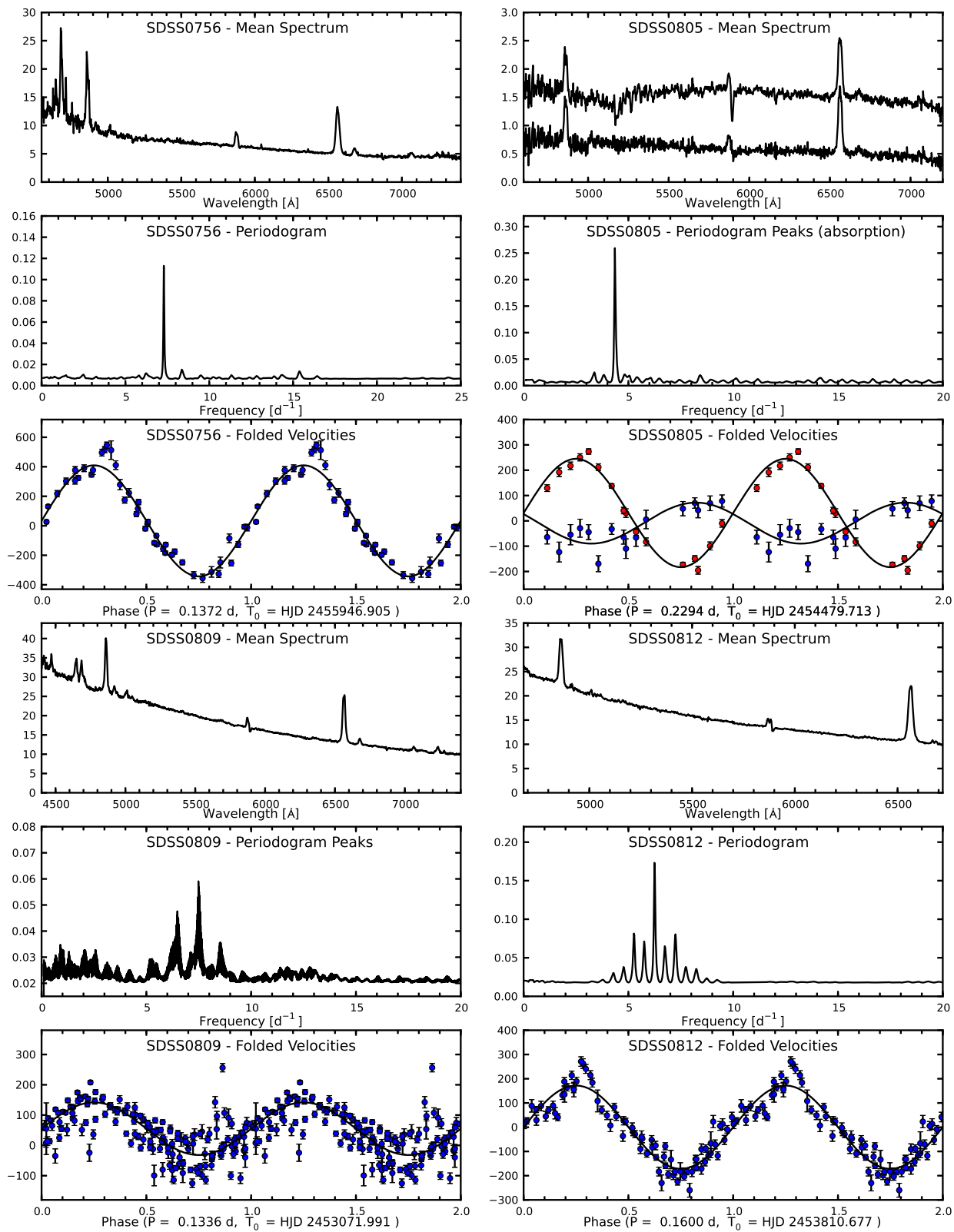}
\vspace{0.4 truein}
\caption{{\it Caption on next page.}}
\label{fig:cvplot1}
\end{figure}

We estimate the secondary's spectral type (see \ref{sec:secondaries})
to be K4 $\pm$ 1.5 subtypes, and that it contributes a flux
corresponding to $V = 19.0 \pm 0.2$ mag 
(see Fig.~\ref{fig:cvplot1}).  The secondary's
absorption-line velocities give a clean determination of the
orbital period, which is 5.51 hours; the H$\alpha$ emission
velocities are not nearly as well-behaved, but 
corroborate the absorption-line period.  
We have 16 velocities from 2008 January, and only a single
point from 2008 March, which happened to fall near 
maximum velocity.  The orbital frequencies that fit
all the data are given approximately by 
$1/P_{\rm orb} = 4.353 \pm 0.021 N \ {\rm cycle\ d^{-1}},$
where $N$ is an integer and $-2 \le N \le +2$.

The secondary star
is a little warmer than typically found in systems in this
period range \citep{kniggedonor}; given this, the evolutionary
scenarios computed by \citet{bk00} suggest that its
mass is $0.75 \pm 0.2$ M$_{\odot}$.  
The \citet{schlegel98} give a total 
reddening $E(B-V) = 0.03$ for this celestial
location ($l = 214.76, b = 19.81$), so the extinction
correction is small.  The Monte Carlo procedure 
described in Sec.~\ref{sec:secondaries} gives 
$2500 \pm 400$ pc for the distance; 
at 2500 pc, the system would be 850 pc from the 
Galactic plane.
     
The emission-line velocities are so noisy that they do
not tell us much about the dynamics of the system, but
the absorption velocities should trace
the secondary's  motion with reasonable fidelity.  The velocity
semi-amplitude, $K = 216 \pm 12$ km s$^{-1}$, is 
fairly large for this orbital period.  In addition, as 
noted above, the secondary's relatively early spectral type may
indicate that it has not lost much mass.  Taken 
together, the $K$-velocity and secondary mass estimate
suggest that the white dwarf is on the 
massive side, and that the orbital inclination is 
not too far from edge-on. 
As an illustration, with $K = 216$ km s$^{-1}$ and
assuming $M_2 = 0.75$ M$_{\odot}$, the system would have
$i = 90$ degrees (edge-on) with $M_1 = 0.85$ M$_{\odot}$;
lower inclinations with these parameters would
require still higher white-dwarf masses.

Our orbital period is corroborated by \citet{woudt12},
who find ellipsoidal variations implying
$P_{\rm orb} = 0.2287(5)$ d.  While they do not detect
eclipses, the ellipsoidal variation also suggests
a fairly high orbital inclination.

\subsection{SDSS J08091+3814 = HS 0805+3822} 

This novalike variable was discovered in the Hamburg
Quasar survey, and rediscovered in the SDSS.  
\citet{rodrig07} found apparent grazing eclipses;
their complete photometric data set was best fit with 
$P_{\rm orb} = 3.22$ hr, but the eclipses alone
could not rule out a period near 3.04 hr; 
\citet{unveils} cite the 3.22-hr period.
\citet{linnell07} modeled the system
using ultraviolet and optical spectra.

Our spectra come from four observing runs, but 
all appear similar, with a strong blue continuum
and Balmer, HeI, and HeII emission lines.  

Our best radial-velocity period, near 3.2068 hr, agrees 
roughly with the
period adopted by \citet{rodrig07} and \citet{unveils},
but disagrees on a finer scale. 
\citet{rodrig07} find 0.1340385(84) d, or 3.2169(2) hr,
based on a count of 313 cycles in between their first
and second grazing eclipse (their Table 1), but our 
radial velocities are not fit well at that period.  
However, an alternate fit assuming 314 cycles
between their first two eclipses gives a best-fit period near
3.2069 h, in excellent agreement with the best choice
of spectroscopic alias.  The 314-cycle eclipse fit
is not quite as good -- while their first two eclipses
fit almost perfectly, their last two miss by $-274$ and
$+267$ s respectively -- but their best fit already shows
deviations nearly this large, so the grazing eclipse
evidently jitters somewhat in phase.  Remarkably, although the
radial velocity periodogram is riddled with aliases
generated by different choices of cycle count between
observing runs, the single best-fitting alias is 
the {\it only} radial-velocity alias that is compatible 
with the 314-cycle eclipse ephemeris.  Because of this 
concordance, we are reasonably sure we have chosen 
the long-term velocity correctly, and quote the period in 
Table~\ref{tab:parameters} with a correspondingly
small uncertainty.

\citet{rodrig07} include this object in a sample of
SW Sex stars. In eclipsing SW Sex stars, the emission-line radial velocity
curves and eclipses are offset in phase, in the sense that
the eclipse occurs somewhat earlier than one would expected
if the emission lines traced the white dwarf's motion.  
Fortunately, our radial velocities and the eclipses observed by 
\citet{rodrig07} were obtained fairly close together in time  --
close enough so that we can compare their phases.  The zero
point of their eclipse ephemeris (their Eqn.~5) occurs
2555.348(44) cycles after our radial-velocity epoch 
(Table~\ref{tab:parameters}).  The radial-velocity
is the red-to-blue crossing of the emission lines, so
if the emission lines traced the white dwarf, the eclipse
would occur at phase $\phi = 0.5$ in the radial velocity
epehmeris.  At $\phi = 0.348(44)$, the eclipse indeed occurs
early, by $55 \pm 16$ degrees of phase.  This is similar to the
phase offset seen in other SW Sex stars. 

SW Sex stars tend to show absorption features in the 
HeI lines that appear at positive velocities around phase 
0.3-0.4, drift toward lower velocity, and then disappear sometime
after phase 0.5.  Fig.~\ref{fig:sdss0809trail} shows a
phase-averaged greyscale representation of our spectra.
The phase-dependent absorption is obvious, and 
behaves exactly as expected for an SW Sex star.  The 
Balmer lines have similar behavior, which is obscured here
by the choice of greyscale.  The 
Mg lines around 5170 also have transient absorption, a
feature seen in PX And as well \citep{thorstensen91}.  
There is also a pair of weak features around $\lambda\lambda$
6348 and 6371, which we tentatively identify as SiII.  
HeII $\lambda$5411 is weakly visible in emission, without
transient absorption. 

\citet{rodrig07} classified this as an SW Sex star without 
benefit of time-series spectroscopy.  The more detailed 
data presented here resoundingly confirm their classification,
as well as refining the orbital period.

\subsection{SDSS J08129+1911} 

\citet{gulsecen13} present three nights of time-series photometry 
of this object that shows eclipses with a period of 0.160046(46) d and
a depth of somewhat over 1 mag.
Our radial velocity period, 0.1600(2) d, is consistent with this but
somewhat less precise.  They also find evidence for a periodicity
near 0.148159 d, which they interpret as a possible negative 
superhump.

The spectrum has a blue continuum and rather modest emission lines --
H$\alpha$ has an equivalent width of about 27 \AA\ and 
a FWHM $\sim 1100$ km s$^{-1}$.  The spectrum, and 
the 3.84 h orbital period, suggests that this is a
novalike variable, or possibly a Z Cam-type dwarf
nova in an extended standstill.  Negative 
superhumps are usually seen in novalikes in this period
range, so the possible negative superhump would tend to 
favor the novalike interpretation.  Also, the CRTSS
light curve shows the source steady, with occasional
downward excursions in which the object was apparently caught
in eclipse (see below), while a Z Cam star would be expected to 
return to quiescence from time to time.
The SDSS spectrum shows only a 
weak bump near HeII $\lambda 4686$, suggesting that this
is not a magnetic system.  The phase-averaged spectrum
(not shown) does not show phase-dependent absorption,
suggesting that this is not an SW Sex star.

Because of the eclipse, we examined the CRTTS magnitudes using
the procedures described for SDSS J07568+0858.   This resulted
in a refined $P_{\rm orb} = 0.1600525(3)$ d.  Fig.~\ref{fig:crtscurves} 
shows the CRTTS data folded on the refined period. 

\subsection{SDSS J08138+2813} 

The mean spectrum of SDSS 0813 shows a strong, blue
continuum; the emission equivalent width of H$\alpha$ is around
40 \AA , and its FWHM is $\sim 950$ km s$^{-1}$.  Emission
at HeII $\lambda$4686 has about half the strength of 
H$\beta$.  \citet{szkodyiv} found a preliminary 
2.7 h radial velocity period, roughly consistent with
our measurement, $P_{\rm orb}$ = 2.928(9) h.  This appears
to be a novalike variable just at 
the long end of the period gap.

As with most of the systems studied here, the flux level
in our spectra is consistent with that from the SDSS
spectra, and with the SDSS magnitudes.  However, in the 
Digitized Sky Survey
red image, taken 1989 Nov 02, 
the object is several magnitudes fainter than at other
epochs.  Like many other novalike variables, this object 
therefore appears to be a VY Scl star -- a cataclysmic that 
dips into low states \citep{honeycuttkafka}.  Consistent
with this, the CRTTS light
curve, which is not densely sampled, mostly shows slow
fading over the past $\sim 8$ years from magnitude 16.5 to
$\sim 18$.  The phase-averaged
spectrum (not shown) is unremarkable, and in particular does not show
phase-dependent absorption.

\subsection{SDSS J08387+4910} 

This is an SU UMa-type dwarf nova, for which \citet{kato09} give
a late superhump period $P_{\rm sh} = 102.92(3)$ min. 
\citet{unveils} estimate an orbital period of 
99.7 min based on the superhump.  \citet{szkodyi} give $g = 19.59$,
but the CRTTS
light curve shows a typical
minimum near 18.2.  Our synthesized $V$ magnitude is 18.5, so the
SDSS magnitude appears to have been taken
during an especially low state. 
The flux level in our spectrum is also significantly brighter than the 
SDSS spectrum \citep{szkodyi}, though the spectral appearance is
fairly similar.  There are differences; in our spectrum, the emission equivalent
width of H$\alpha$ is $\sim 110$ \AA, about half that in the 
SDSS spectrum, and its FWHM is $\sim 1600$ km s$^{-1}$.
H$\alpha$ is barely double-peaked, suggesting an 
intermediate orbital inclination.   

The radial velocities indicate $P_{\rm orb} = 101.29(0.28)$ min,
slightly longer than predicted based on the superhump,
and yields a superhump period excess 
$$\epsilon = {P_{\rm sh} - P_{\rm orb} \over P_{\rm orb}} = 0.016(3).$$
\citet{kato09} give a relation (their eqn.~4) between the late
superhump period (their $P_2$) and the superhump period excess,
which predicts $\epsilon = 0.027$ at this period.  The 
superhump period excess $\epsilon$ can be used as a measure of 
the mass ratio $q = M_2 / M_1$ \citet{pattlate}.  If confirmed,
the unexpectedly low $\epsilon$ should indicate an atypically
small $q$.  The spectroscopic period is based on only two nights
of data, though, so a more precise measurement would be especially
desirable.

\subsection{SDSS J08440+0239 = V495 Hya} 

This is listed as a U Gem-type dwarf nova in the
General Catalogue of Variable Stars \citep{gcvs}\footnote{Updated
versions of the GCVS can be retrieved from 
http://www.sai.msu.su/gcvs/gcvs/iii/html/},
with an uncertain magnitude of 14 at maximum.
The CRTTS light curve shows two outbursts in 
the five seasons with good coverage.

Of our 92 spectra, 89 are are from 2006 January, and
the other three are from 2004 January.  The spectra
from the two runs appeared similar, but the sparse 
2004 January spectra were not used in the analysis.  

The mean spectrum shows relatively narrow Balmer lines, 
with H$\alpha$ having a FWHM of 600 km s$^{-1}$.
HeII $\lambda$4686 is barely detected, and the FeI 
line near $\lambda$5169 is present as well.  The continuum
is nearly flat, but rises slightly to the red and shows
the undulations from a weak M-dwarf contribution.
The velocity time series from 2006 January spans
8.7 hours of hour angle, and gives an unambiguous
period near 4.97 hours, among the longer periods
found here.

Using the procedure described in Section~\ref{sec:secondaries},
we estimated the secondary's spectral type to be 
M2.5 $\pm$ 1.5 subclasses, 
and its flux contribution to be equivalent to 
$V = 21.5 \pm 0.7$.  
Repeating the procedure using the 
archived SDSS spectrum and M-dwarf templates assembled
from SDSS data
by \citet{bochanskitemplate} gave an almost
identical result.
The secondary's spectral type is close to the that
expected at this orbital period \citep{kniggedonor}.
The \citet{bk00} evolutionary calculations indicate
a wide range of possible secondary masses at this
$P_{\rm orb}$, so that the secondary's
radius is constrained only to $-5 \log_{10} (R / R_{\odot}) = 1.6 \pm 0.4$.
In the observed spectral type range, a 1 R$_{\odot}$ star 
would have $M_V = 8.5 \pm 1.0$.  The absolute magnitude
of the secondary is therefore $M_V = 10.1 \pm 1.1$, 
yielding $m - M = 10.9 \pm 1.3$.  The \citet{schlegel98} dust
maps give a small reddening of $E(B-V) = 0.04$ at this location 
($l = 224.08, b = 26.13$), so $(m - M)_0 = 10.8 \pm 1.3$.
The best nominal distance is therefore 1400 pc, but with almost
a factor of two uncertainty in either direction.
The calculation above, combined with 
the observed magnitude, suggests that this system has $M_V \sim 8$,
consistent with its identification as a dwarf nova.

\subsection{SDSS J09127+6209} 

This system was discovered both by SDSS \citep{szkodyviii} and apparently
independently by \citet{wils2010}, who searched SDSS and other catalogs for
CV candidates.  In our average spectrum, H$\alpha$ has an emission
equivalent width of nearly 200 \AA, double-peaked profile with peaks
at $\pm 500$ km s$^{-1}$, and and a FWHM of 1800 km s$^{-1}$.  The
double peaks and sizeable radial velocity amplitude ($K = 100 \pm 15$ km s$^{-1}$)
suggest that the inclination is fairly high, but the system is not known 
to eclipse.  Our spectra, nearly all from 2013 January, do not extend
far enough to the red to detect the red dwarf that features prominently in 
the \citet{wils2010} spectrum.  

Only the 2013 data were extensive enough to analyze for radial velocity.
The parameters listed in Table~\ref{tab:parameters} are from 
the 2013 January data.  The period, 115.39(13) min, agrees with the
$\sim 113$ min preliminary period found by \citet{szkodyvii}.  The spectrum
and period are typical of SU UMa-type dwarf novae.  The CRTTS shows a single
possible outburst of $\sim 2$ mag, but coverage is not heavy.
If a future outburst shows superhumps, our period is precise enough to 
yield an accurate superhump period excess.

\subsection{SDSS J09168+2849 = HH Cnc} 

This dwarf nova shows four outbursts in its
CRTTS light curve, which covers nine seasons, so outbursts
are fairly infrequent.  The spectrum shows a contribution
from an M-type companion.  We assume this is
from the secondary star, but the M-type spectrum is too faint to show
radial velocity variation, so we cannot entirely rule out
a superposition.  The FWHM of the H$\alpha$ emission is
490 km s$^{-1}$ in our summed spectrum, which is fairly narrow
for a CV.  

As one might expect from the narrowness of the line, the 
radial velocity amplitude is low, and most of the time the
orbital motion is masked by small fluctuations
probably due to small changes in the emission-line profiles.  
Our efforts to find $P_{\rm orb}$ were stymied on several
occasions, but in 2013 February and March the velocities 
varied smoothly.  The period proved to be 
$P_{\rm orb} = 4.43 \pm 0.01$ hr, without ambiguity. 
The velocity data used in Fig.~3 are from the 
2013 February/March observing run.

Using the secondary's contribution, we can estimate a
distance in the manner outlined in Sect.~\ref{sec:secondaries}.
The M dwarf appears to be in the range M1.5 to M4.0.
A rather wide range of masses is possible under different
evolutionary scenarios; for purposes of computing the 
M-dwarf radius, we assumed a uniform distribution
from 0.15 to 0.50 solar masses.  The Monte Carlo
simulation gave a distance of 
$1500 \pm 500$ pc.  At 1500 pc and $b = 43^{\circ}$, 
HH Cnc would lie 1.0 kpc from the Galactic plane.  

\subsection{SDSS J09197+0857 = LV Cnc} 

Both our mean spectrum and the discovery spectrum \citep{szkodyiv} 
show prominent absorption wings in the Balmer lines, which 
arise from a white-dwarf photosphere; \citet{szkodyiv} also estimated
a $\sim 1.4$ hr period from radial velocities. 
\citet{dillon08} found a photometric period of $81.6 \pm 1.2$ min in a 
single night of time series photometry.   \citet{mukadam07}  
detected a 260 s oscillation in the light curve, which they interpreted
as a likely non-radial oscillation, an interpretation corroborated
by \citet{woudt12}.  The radial-velocity 
period found here, 81.32(17) min or 1.36 hours, agrees with and
improves on the earlier determinations. 
The prominent 
white dwarf photosphere, very short $P_{\rm orb}$, and 
apparent white dwarf oscillations are all reminiscent of 
GW Lib \citep{szkodygw, vanzyl04, thorshortest}.  The
CRTTS shows no clear-cut outbursts despite heavy coverage over eight years, 
though the source grew brighter by $\sim 0.5$ mag toward the end of the 
2006 season.

The USNO B1.0 catalog \citep{usnob} lists a proper motion of 60 mas yr$^{-1}$
in a position angle of 262 degrees.  The sizeable proper motion, as 
well as the visibility of the white dwarf, suggest a relatively nearby
distance.  We therefore began parallax observations 
\citep{thorparallax1, thorparallax2} of this object
in 2006 January, and now have 78 images taken at 12 epochs. 
This sequence yields a relative parallax $\pi_{\rm rel} = 4.7 \pm 1.7$ mas, and 
a relative proper motion of 59 mas yr$^{-1}$ in a position angle
of 266 degrees, in excellent agreement with the USNO B1.0.  Correcting
for the estimated distances of the reference stars gives $\pi_{\rm abs} = 
5.5$ mas.  The uncertainty estimated from the scatter of the 
measured parallaxes of similarly faint stars in the field corroborates
the uncertainty derived from the goodness of fit, so we adopt
$\sigma_{\pi} = 1.7$ mas.  Since $\pi_{\rm abs} / \sigma_{\pi} < 4$, 
the classic Lutz-Kelker correction does not formally converge, so following
\citet{thorparallax1} we adopt a Bayesian approach to the distance
estimate, which combines the parallax with prior information
(``priors'') from the proper motion, and estimated luminosity.  
We use GW Lib to construct the luminosity prior.  \citet{thorparallax1}
gives a distance of 104 ($30,-20$) pc for this star, and our
parallax images give a mean $V = 16.96$, while SDSS 0919 has 
$V = 18.30$, which (if it were identical to GW Lib) would put 
it about 190 pc distant with $M_V \sim 11.9$.  Taking this as
our luminosity prior, assigning it a 1.5 mag uncertainty, then
yields a final Bayesian estimate of 195 $(+54, -39)$ pc -- almost
exactly the inverse of the parallax, and also just as expected
from scaling the GW Lib distance.  

At 195 pc, the transverse velocity (corrected to the local 
standard of rest, or LSR) would be 43 km s$^{-1}$.  The velocity fit
in Table \ref{tab:parameters} gives a barycentric systemic velocity of $+73$ km s$^{-1}$; taking
this at face value, the space velocity relative to the LSR 
is ${U,V,W} = {+71,-30,+11}$ km s$^{-1}$.  This is a rather
large in-plane velocity, suggesting that the system belongs to 
the old disk. 

\subsection{SDSS J09224+3307} 

The mean spectrum is typical of a dwarf nova at minimum light;
H$\alpha$ has an emission equivalent width of 160 \AA\ and a
FWHM of 1300 km s$^{-1}$.  HeI $\lambda$5876 is double-peaked,
with the peaks separated by 880 km s$^{-1}$, but the double
peaks of the Balmer lines are less distinct.  The orbital inclination
is likely to be unremarkable.

We observed SDSS0922 with the WHT on three nights in 2007 March, under
mediocre conditions, and found a $\sim 90$ min radial velocity
period, but the daily cycle count remained ambiguous.
MDM observations taken 2013 February covered a 
greater range of hour angle and resolved the ambiguity.
Table \ref{tab:parameters} gives the best fits to both
data sets and the weighted average period, which amounts to
94.56(12) min.  The two data
sets give significantly different values of the average
velocity $\gamma$, but we do not think this discrepancy is 
likely to be physically meaningful.  To construct the periodogram  
and folded velocity curve (Fig.~\ref{fig:cvplot3}),
we adjusted the data to their weighted average $\gamma$ velocity
by subtracting 24 km s$^{-1}$ from the WHT velocities, and adding
33 km s$^{-1}$ to the MDM velocities.

The period and spectrum both suggest that this is a dwarf nova
of the SU UMa subtype, but outbursts have not been confirmed.
The CRTTS light curve shows variation mostly between 18 and 19 mag,
with a few points brighter than 17th mag that may have been weak
or poorly-covered outbursts.

\subsection{SDSS J09459+2922} 
The mean spectrum shows a blue continuum with strong emission lines --
H$\alpha$ is single-peaked, with an emission equivalent width of 180 \AA\
and a FWHM of 1000 km s$^{-1}$. 
Although the system is relatively faint at $g = 19.1$, 
the H$\alpha$ emission velocities were well-behaved, making the
period determination relatively straightforward.  We have only 
22 velocities from two successive nights, yielding $P_{\rm orb} = 92.0
\pm 0.3$ min. The velocity amplitude $K = 255 \pm 22$ km s$^{-1}$, 
which is larger than usually seen in disk systems; in addition,
emission line profiles are single-peaked, while an edge-on disk system tends
to have doubled lines.  This 
suggests that the system is magnetic, but the HeII $\lambda$4686 line,
which is usually strong in magnetic systems, is quite weak.  
Time-series polarimetry would be a useful diagnostic.

The CRTS light curve mostly shows variation between 18.5 and 19.5 
with occasional points fainter than 20th, and no outbursts.
 
%
%
%
\subsection{SDSS J10133+4558} 

This did not yield a period easily, so we observed it in
in 2008 January, 2010 March, and 2011 January.  
On the last run, we used the 1024$^2$ CCD `Templeton', which 
covered a narrower spectral range than the `Echelle' chip used
on the other runs.  The mean spectrum from the earlier runs
gives a synthesized $V$ = 19.0, but the 2011 January data
give $V$ = 19.6, as a consequence of a fainter continuum.
The emission line fluxes did not appear to change significantly.

The number of cycles elapsed between the widely separated observing runs
is uncertain.  Because of this, the period give in Table~\ref{tab:parameters}
is the weighted average derived from separate fits to the 
2008 January and 2011 January velocities, which were 
extensive enough to fit.  

The spectrum (Fig. \ref{fig:cvplot4}) is typical of a dwarf
nova at minimum light, but the CRTTS light curve shows no
outbursts despite 232 epochs of coverage in 8 seasons, so
outbursts are likely to be infrequent.  Infrequent outbursters
often resemble WZ Sge, for which all observed outbursts have
been superoutbursts (see, e.g. \citealt{pattwzsge}).  It remains to be seen if this object
behaves similarly.

\subsection{SDSS J10197+3357 = HS 1016+3412 = AC LMi}

This object was originally found in the Hamburg Quasar
Survey; \citet{aung06} classified it as a dwarf nova.  
The CRTTS light curve shows the source mostly near 18th mag,
with two outbursts to $\sim 15$ mag in eight years.  
\citet{aung06}
measured $P_{\rm orb} = 114.3 \pm 2.6$ min from emission-line
radial velocities.  The period we find here, 
$114.0 \pm 0.2$ min, confirms and refines their result. 
Dwarf novae in this period range are generally SU UMa 
stars, which show superoutbursts during which they develop
superhumps.  If this proves to be the case for this object,  
a usefully precise superhump period excess can be calculated
using our improved period.

\subsection{SDSS J10233+4405 = NSV 4838 = UMa8} 

\citet{downescat} list this SU UMa-type dwarf nova as `UMa8'.
\citet{imada09} studied two superoutbursts
and found a superhump period $P_{\rm sh}$ near 0.0699 d, or 
100.6 minutes; comparing this to our preliminary
orbital period, they estimated a mass ratio $q = 0.13$.  

Our spectrum (Fig. \ref{fig:cvplot4}) is typical of a quiescent
dwarf nova.  The H$\alpha$ emission line has an
EW of 140 \AA , 
two peaks separated by $\sim 900$ km s$^{-1}$, and
a FWHM $\sim 2100$ km s$^{-1}$, suggesting
a high orbital inclination.  We cannot determine an 
unambiguous daily cycle count from our radial velocities --
the most likely choice is 14.7 cycle d$^{-1}$, but
15.7 and (oddly) 10.4 cycles d$^{-1}$ are possible.
However, the single best choice, corresponding to 
97.9(8) min, is as expected given the superhump period,
so the alias choice is secure.  Because our data are from several
observing runs, with an unknown cycle count between them, 
we derived the period in Table~\ref{tab:parameters} by 
fitting the velocities from different runs separately
and averaging the results. 

\subsection{SDSS J10280+2148 = 1H 1025+220 = Leo5}

This object was apparently discovered because of its X-ray emission,
and listed as `Leo5' in the \citet{downescat} catalog, which credits 
the identification to R. Remillard.  The spectrum published by 
\citet{munari97}, the SDSS spectrum \citep{szkodyviii}, and our
averaged spectrum (Fig. \ref{fig:cvplot4}) are all similar,
showing a blue continuum with weak Balmer emission lines.  Although the 
spectrum suggests this is a novalike variable, \citet{wils11} studied
its variation and found it to be a Z Cam star.  

Most of our spectra are from 1998 January, but we have data from
several other runs up as well\footnote{The data were published in 
author CJT's Ph.D. thesis \citep{taylorphd}, but have not been published in a journal
article until now}.  The radial velocity period, 
$3.505(4)$ hr, is determined with little ambiguity in the daily 
cycle count, but the data do not select the number of cycles
elapsed between observing runs.  The quoted period uncertainty
here and in Table~\ref{tab:parameters}
is derived from the 1998 January data alone.  The period is typical for a
Z Cam star.

\subsection{SDSS J11313+4322 = MR UMa}

This CV was discovered by \citet{wei97} as the optical counterpart of
a ROSAT X-ray source.  It is an SU UMa star, for which \citet{kato09} 
list a superhump period near 0.0652 d in the main part of the 
outburst.  \citet{szkodyv} list $g = 16.17$, but our spectrum
indicates $V = 17.9$, so the star was evidently above minimum when the 
SDSS direct image was taken.  The continuum level in the 
SDSS spectrum (in Fig.~1 of \citealt{szkodyv}) is nearly identical to ours, so our
spectra were apparently both taken in quiescence. 


\subsection{SDSS J11322+6249} 
The mean spectrum shows a strong, blue continuum and single-peaked
emission lines.  The H$\alpha$ emission line has an equivalent width of 
85 \AA\ and a FWHM of 800 km s$^{-1}$.  H$\beta$ is stronger
than H$\alpha$, with a flux ratio $F({\rm H}\beta) / F({\rm H}\alpha) \sim 1.7$.
HeII $\lambda$4686 is present at $\sim 1/7$ the strength of H$\beta$.
We took spectra of SDSS1132 on two observing runs, in 2008 January and
2008 March.  The January data indicated a period near 99 min, and the 
March data resolved ambiguities in the daily cycle count.  
The number of cycles that elapsed in the 
interval between the two runs is unknown, but the set of 
possible periods can be expressed as 
$$P = {45.228 \pm 0.004\ {\rm d} \over 656 \pm 7},$$
where the denominator is an integer.  

The subclassification of this object is ambiguous.  The 
continuum and HeII line suggests that this might be an 
AM Her star (or polar), but the line profiles remain 
symmetrical around the orbit, while AM Her stars
typically show highly variable, asymmetric profiles from the 
accretion column (see, e.g., \citealt{schwope99}).  
The object is listed in the ROSAT
faint source catalog \citep{vogesrosatfaint}.  Dwarf nova
outbursts do not appear to have been detected; the CRTTS 
light curve shows erratic variation between magnitude
17.5 and 19.5 in their system.  This may be 
a novalike variable with a period shortward of the 
gap, which if confirmed would be unusual.

\subsection{SDSS J12442+3004} 

This dwarf nova was discovered by the CRTTS prior to its
listing by SDSS; it is also listed in the GALEX catalog and
was selected as a QSO candidate by \citet{atlee07}. 
The mean spectrum shows the strong emission lines typical
of short-period dwarf novae at minimum light.  We have
spectra from two observing runs, one in 2009 March and the
other in 2010 March.  The 111.5-min period is found from 
an unambiguous daily cycle count but the number of 
cycles elapsed between observing
runs is unknown; the exact period is constrained to 
$$P_{\rm orb} = {382.824 \pm 0.002\ {\rm d} \over 4944 \pm 25},$$
where the denominator is an integer.  We are unaware of 
any detection of superhumps, but at this orbital period
they are likely to occur.

\subsection{SDSS J14299+4145} 

The SDSS spectrum \citep{szkodyiv} shows a blue continuum
with single-peaked emission lines, and with HeII $\lambda4686$ 
about half the strength of H$\beta$.
The Balmer decrement is inverted, with H$\beta$ much stronger than 
H$\alpha$.  Our mean spectrum is similar.  We have data from 
2005 March and 2005 June; in March, our spectra implied
$V = 18.2$, but in June it had faded to $V = 19.2$.  The 
object appears much fainter on all the digitized Schmidt plate 
survey images than it does in SDSS, so it is evidently highly
variable.  The CRTTS data are very sparse in this location, but
show variation between magnitude 18 and 21.  

Our H$\alpha$ radial velocities favor a period near 
99 min, or 14.61 cycle d$^{-1}$, but daily cycle-count
aliases near 15.68 and 13.57 cycle d$^{-1}$ are also
possible.  The number of cycles elapsed in the 100-day
interval between our 2005 March and June runs is also
not determined.  Our period corroborates the $\sim 1.4$ hr 
candidate period found by \citet{szkodyiv} in 2.5 hr of spectroscopy,
but neither determination is definitive.

If the candidate spectroscopic 
period is $P_{\rm orb}$, as is nearly always the case,
then $P_{\rm orb}$ is unusually short given the star's other characteristics.
The emissions at HeII $\lambda4686$ is about half the 
strength of H$\beta$, which is typical for intermediate
polars (IPs, or DQ Her stars) -- systems containing 
a magnetic white dwarf that rotates more quickly than
the binary orbital frequency.  Most IPs, however, have
longer orbital periods than this.    
Polars, or AM Her stars -- systems in which a magnetic
white dwarf rotates synchronously -- often have periods
as short as SDSS1429, and can also spend long intervals 
in very low states, as this object evidently does.  
However, most polars have 
$\lambda 4686$ roughly equal to that of H$\beta$, and the 
velocity amplitude tends to be very large, because of the
high infall velocities in the accretion funnel close to the white dwarf.  
One of the spectra obtained by \citet{szkodyiv} showed double-peaked
Balmer lines and HeII $\lambda$4686 approximately equal to H$\beta$.
The double-peaked lines might be due to transient absorption, which 
would be reminiscent of an SW Sex star, but as \citet{szkodyiv} remark,
SW Sex phenomena tend to be found in objects with $P_{\rm orb} > 3$ hr.

The nature of this object therefore remains rather problematical.
If it is a magnetic system, time-series photometry might
reveal a white dwarf pulsation period.

\subsection{SDSS J15046+0847 = ``Boo1''} 

\citet{fili85} discovered this object on an objective-prism
plate.  The spectrum appears typical for a dwarf nova at minimum light,
but outbursts do not appear to have been recorded; in particular, the
CRTTS shows no outbursts despite very good coverage.  The emission
lines are single-peaked, and the FWHM of H$\alpha$ is 770 km s$^{-1}$,
suggesting a fairly low orbital inclination.
Our spectra, from three successive nights in 2010 March, span only
5.3 hours of hour angle, but the radial velocity period is determined
to be 116.5(3) min without significant ambiguity.  This period
is typical of SU UMa-type dwarf novae, so superhumps are likely
in the event of an outburst.

\subsection{SDSS J15365+3328} 

This object was discovered independently by the CRTTS as a dwarf
nova.  The CRTTS light curve shows four outbursts, 
reaching an apparent magnitude $m \sim 15$, with a 
minimum typically $m \sim 18.5$, with some fainter 
interludes.  

The mean spectrum is typical of a dwarf nova at minimum light,
with emission lines of H, HeI, and FeII $\lambda$5169.
The lines are slightly double-peaked, with the red and
blue peaks of H$\alpha$ separated by $\sim$600 km s$^{-1}$.
The H$\alpha$ line has an emission equivalent width near 
100 \AA\ and a FWHM of $\sim$1500 km s$^{-1}$.  The period,
132.7(3) min, is a unusual for being near the lower edge 
of the gap, but it appears to be reliable; Monte Carol tests give a 
probability of less than 1 per cent 
for a cycle count error.  At this $P_{\rm orb}$,  the superhumps and
superoutbursts characteristic of an SU UMa star should occur.


\subsection{SDSS J15567$-$0009 = V493 Ser} 

\citet{woudt04} reported photometry and found a period 
of 0.07408(1) d; their choice of daily cycle count was, however,
not entirely secure. 
\citet{kato09} observed a superoutburst in 
2007 that showed a mean superhump period of 0.082853(5) d.
Because superhump periods are almost always only a few per cent
longer than orbital periods, they concluded that the 
\citet{woudt04} period is the one-day alias of the correct
period, which would then be near 0.08001(1) d.  Our
radial-velocity period supports this revision.  At the
time of our observations (2003 June 20-24), the source transited 
well before midnight, so we obtained only a 5.3-hour span of
hour angle.  Even so, the Monte Carlo test 
\citep{tf85} gives a discriminatory power greater than 0.9,
and a correctness likelihood near unity for our 
alias choice.  The radial-velocity period is 0.08012(8) d
-- consistent, within the errors, with that inferred by
\citet{kato09}.  The alias choice is therefore 
secure, and the \citet{woudt04} period is superseded.

\subsection{SDSS J16101+0352 = V519 Ser} 

This object was discovered as the counterpart of a ROSAT X-ray
source \citep{jiang2000}.  \citet{rodrigues06} detected strong circular 
polarization, confirming it as a polar, or AM Her star.  
From their photometry and polarization they found 
$P_{\rm orb} = 0.13232(4)$ d.  

Our radial velocities span the interval from 2000 April to 
2011 June.  An unambiguous cycle count over that
interval yields $P_{\rm orb} = 0.1322739(2)$ d, 
consistent with the \citet{rodrigues06} result but 
more precise because of the long time base. 

The discovery spectrum shows a clear contribution from an M-dwarf 
secondary star, which \citet{jiang2000} classify as M4 or M5. 
This is not evident in SDSS spectrum \citep{szkodyvii}, which has 
$\sim 8$ times the flux, obscuring the secondary's contribution.
Our mean spectrum (Fig. \ref{fig:cvplot6}) shows the source somewhat brighter
than the discovery spectrum, and the secondary is detected. 

To classify the secondary, we averaged together the spectra
with the lowest continuum levels (fainter than synthesized
$V = 18.3$), and compared this average against a library of 
spectra of M dwarfs taken with the same instrument.  Types
M4 through M5 did appear to match somewhat better than earlier
types.  An argument similar to that used in SDSS J08055+0720 
-- using the secondary's inferred surface brightness and the
constraints on its radius given the orbital period -- 
gives a distance estimate of 450 (+120,$-$100) pc for this
system.

\subsection{SDSS J16191+1351} 

The CRTTS light curve of this object shows irregular
variations between $m = 17$ and 18 magnitude, punctuated
by a large number of short outbursts to $m \sim 15$, 
marking this as a frequently-outbursting dwarf nova.  
Indeed, the source was in outburst during follow-up 
spectroscopy reported by \citet{szkodyvii}.  

The continuum in our spectrum sweeps upward toward the 
blue, suggesting that the object was not completely at
minimum light.  The synthetic $V = 17.24$ corroborates
this conclusion.  Even so, a weak contribution from a
secondary star is discernible, in addition to the emission
lines typical of dwarf novae.  Using the procedures
described in Section~\ref{sec:secondaries}, we classify the secondary
as K4 $\pm$ 2 subclasses, and estimate a distance
of 3100(+800,$-$700) pc.  The Galactic latitude of 
nearly 40 degrees, so our distance puts the system about
2 kpc away from the Galactic Plane.

Both the emission and absorption line velocities show
orbital modulation at the same period (within the uncertainties).
The difference in the phase of the emission and absorption
modulations is 0.478(16) cycles, consistent with the 
value of 0.5 expected if the emission traces the
motion of the white dwarf\footnote{The mean velocities $\gamma$
of the absorption and emission -- 79(7) and 19(8) km s$^{-1}$
respectively -- are formally very different, but the 
emission velocities especially can be skewed by line-profile
effects, so we are reluctant to ascribe physical 
significance to this.}.  For plausible masses, the velocity 
amplitude $K$ suggest an intermediate orbital inclination.
If we assume that (1) the emission and absorption $K$-velocities 
faithfully represent with the white dwarf and secondary
motions respectively, and (2) the white dwarf has $0.7$ M$_{\odot}$, 
then the nominal $K_2 / K_1$ gives 0.56 M$_{\odot}$ for the 
secondary, and the inclination is around 40 degrees.  
Note that we have not measured this -- the calculation is 
simply to show that the data are consistent with ordinary
values, and that the system is rather unlikely to eclipse.
  
\subsection{SDSS J16539+2010 = V1227 Her} 

This is an SU UMa-type dwarf nova.  When \citet{szkodyv} 
performed follow-up photometry on this object, they found it 
in outburst and discovered a 1.58-hour superhump modulation,
implying an orbital period near 91 minutes.  \citet{kato10}
report $P_{\rm sh} = 93.54(4)$ min early in the outburst,
lengthening to 93.91(4) min later.  Our spectroscopic
orbital period is 90.86(13) min, so that $\epsilon = 
(P_{\rm sh} - P_{\rm orb}) / P_{\rm orb}$ evolves
from 0.029 to 0.034 during the outburst.  This is 
approximately as expected at this orbital period (e.g., 
eqn.~2 of \citealt{unveils}).

\subsection{SDSS J16569+2121 = V1229 Her} 

Our spectra are all from 2006 June.
The mean spectrum (Fig.~\ref{fig:cvplot7}) shows broad emission lines with incipient 
double peaks; H$\alpha$ has a FWHM of over 1700 km s$^{-1}$
and an equivalent width $\sim 130$ \AA . Judging from the breadth 
of the lines, the orbital inclination is likely to be 
high. However, \citet{southworth07} did not detected any 
eclipses in a photometric time series longer than the 96.9-min
orbital period found here.  The period and
spectrum suggest that this is
an SU UMa-type dwarf nova; if superhumps and
superoutbursts are detected, this would be confirmed.
The CRTTS light curve shows no outbursts, despite rather
dense sampling, so outbursts appear to be infrequent.

\subsection{SDSS J16598+1927} 

The mean spectrum (Fig.~\ref{fig:cvplot7}) shows a strong, blue continuum and relatively
weak lines -- the equivalent width of H$\alpha$ is only 18 \AA .
The HeI lines are barely detected at this signal-to-noise
ratio, but HeII $\lambda$ 4686 is present.  There is sharp
absorption at the NaD lines, which is very likely interstellar.

\citet{southworth08} obtained a short spectroscopic time series,
which failed to show an orbital period.  Our velocities were
noisy -- we needed 106 velocities spread over five different
observing runs to secure the daily cycle count, so their
non-detection of $P_{\rm orb}$ is not surprising. The 
periodogram of our velocities shows much fine-scale structure near 
the orbital frequency (7.07 cycle d$^{-1}$, or $P_{\rm orb} 
\sim 0.141$ d).  

The spectrum and orbital period of this object are consistent
with a novalike variable, though it could
also be a Z Cam star in an extended high state.
The CRTTS light curve shows only irregular variations 
between 16.5 and 17.5 magnitude, which suggests that this is a
novalike. 

\subsection{SDSS J17301+6247} 

This is an SU UMa-type dwarf nova, with a $P_{\rm sh}$ near 0.0794 d, 
or 114.3 min \citep{kato09}.  The emission lines (Fig.~\ref{fig:cvplot8})
are single-peaked, suggesting
a fairly low orbital inclination.  Combining our  measured orbital period,
110.3(2) minutes, with $P_{\rm sh}$ gives a fractional superhump 
period excess $\epsilon = (P_{\rm sh} - P_{\rm orb}) / P_{\rm orb} = 0.037$,
which is close to the expected value for this $P_{\rm orb}$.

\section{Discussion}

As \citet{unveils} emphasize, the SDSS has brought us 
significantly closer to a true accounting of the CV period 
distribution;
in particular, over 30 per cent of their sample of 92 CVs
discovered in the SDSS
had periods lying in a `spike' in the distribution, which they
defined as being from roughly 80 to 86 minutes.  
The shortest-period object in the present sample is LV Cnc (=SDSS J09197+0857), at 
81.3 min; the next-shortest is V1227 Her (= SDSS J16539+2010), at 90.9 minutes.
LV Cnc is therefore the only object in the `spike'. 
The white dwarf in LV Cnc is readily apparent in the 
spectrum, which marks it 
as one of the least luminous objects in the sample.  
Evidently, the present sample cannot be taken as
representative, because it is skewed toward higher luminosities.
This is not surprising, since spectroscopy of fainter 
targets can be very difficult.
It is fortunate, then, that many of the shortest-period
CVs undergo superoutbusts and show superhumps, which can
be used to derive approximate periods that are good enough
for many purposes (see, e.g.. \citealt{kato09, kato10}).  

In general, longer-period
systems tend to be more luminous \citep{patterson84}, 
and hence more amenable to spectroscopy.  
Indeed, eleven of the 29 systems discussed here 
have periods longer than three hours -- in 
order of increasing period, they are  
SDSS0813, 1610, 0809, 0756, 1659, 1028, 0812,
0916, 0844, 0805, and 1619.  As noted above,
it is relatively easy to find rough orbital
periods for short-period dwarf novae that show superoutbursts.
Spectroscopic studies such as the present one are needed
in part to ensure that the distribution of known CV periods
does not become skewed toward the short end.

The number of known and suspected CVs has exploded in recent 
years because of deep, all-sky variability surveys
such as the CRTTS, ASAS-SN\citep{shappee14}, and MASTER
\citep{lipunovmaster}.  However, because of the 
enormous number of variable stars in the sky, only objects
with large-amplitude variability tend to be flagged
as CVs; in the CRTTS, for example, only objects
with $\Delta m > 2$ mag are flagged as CV candidates
 \citep{drakecrtts}.  The sample of CRTTS CVs is 
consequently biased toward large outburst amplitudes
\citep{thorcrts}.  Intriguingly, though, 
\citet{breedt14}, reviewing more than 1000 CRTTS dwarf
nova discoveries over the past eight years, find that 
the rate at which CRTTS discovers {\it frequently outbursting}
objects has slowed dramatically, while the rate at
which it is finding objects that show no previous
outbursts has {\it not} slowed significantly.
This suggests that there is a sizable population of 
quiescent dwarf novae -- or, more properly, dwarf novae
with very long outburst recurrence times -- waiting to 
be discovered.

The SDSS CV sample, which was selected by color and spectrum,
supports this idea.  Seven of the objects in the present sample  
(SDSS0805, 0919, 0945, 1013, 1132, 1504, and 1656)  
have spectra consistent with dwarf novae at minimum light, 
but without known outbursts, even though some have been
covered extensively by CRTTS.  
Variability surveys have also tended to miss 
other classes of CVs that are not dwarf novae -- the 
so-called novalike variables.  The present sample has
some seven of these, namely SDSS0756, 0809, 0812, 
0813, 1028 (which may be a Z Cam-type dwarf nova), 
1429, and 1659\footnote{Several of these were discovered by
other surveys, but SDSS did select them by color and
obtain spectra.}.     

It's likely, then, that as new CVs are discovered, they
will mostly be (a) dwarf novae that seldom outburst --
possibly with arbitrarily long recurrence times -- and
(b) various kinds of novalike variables.  Since essentially
{\it all} cataclysmic binaries that actively transfer matter
are variable stars, these might be found by looking for
lower-amplitude variables in surveys such as CRTTS, and 
sifting these for unusual colors.  In the more distant
future, a careful analysis of the Large Synoptic Survey 
Telescope\footnote{An up-to-date description of the 
LSST project can be found at http://arxiv.org/abs/0805.2366.}
data should find essentially every CV in the 
southern sky to an impressive magnitude limit.


For convenience, 
we list here some noteworthy results on individual objects:
\begin{enumerate}
\item SDSS J07568+0858 and SDSS J08129+1911 were both previously known
to be eclipsing novalike variables; we refined their eclipse periods using
archival data from CRTTS.  This technique may be of interest to 
other workers in the field.
\item We improve the period of SDSS J08091+3814, and use 
phase-averaged spectroscopy to corroborate its suggested
membership in the SW Sextantis class 
\citep{rodrig07}.
\item We measure a parallax-based distance of 195 (+54, $-$39) pc for the 
shortest-period object in our sample, LV Cnc (SDSS J09197+0857).
This system appears to be a near-twin of the seldom-outbursting
dwarf nova GW Lib.  
\item SDSS J09459+2922 has some characteristics of a magnetic CV, and may 
be a rewarding target for time series polarimetry.
\item The subtype of SDSS J11324+6249 is ambiguous; it may be a
novalike variable with an atypically short period.
\end{enumerate}

\section{Acknowledgements}

We gratefully acknowledge support from NSF grants 
AST-0307413, AST-0708810, and AST-1008217.  
We thank the MDM staff for observing 
assistance, and S\'ebastien L\'epine for taking some of the 
parallax data on SDSS 0919.  

This paper uses data from the
Catalina Sky Survey and Catalina Real Time Survey; the CSS is 
funded by the National Aeronautics and Space
Administration under Grant No. NNG05GF22G issued through the Science
Mission Directorate Near-Earth Objects Observations Program.  The CRTS
survey is supported by the U.S.~National Science Foundation under
grants AST-0909182.

\clearpage

\begin{deluxetable}{lrrrrrl}
\tablewidth{0pt}
\tablecolumns{7}
\tablecaption{List of Objects}
\tablehead{
\colhead{Abbreviation} &
\colhead{Name} & 
\colhead{$\alpha_{\rm 2000}$} &
\colhead{$\delta_{\rm 2000}$} &
\colhead{$g$} &
\colhead{Synth. $V$} &
\colhead{References} \\
\colhead{} &
\colhead{} &
\colhead{[h m s]} &
\colhead{[$^{\circ}$ $'$ $''$]} &
\colhead{(mag)} &
\colhead{(mag)} &
\colhead{} \\
}
\startdata
SDSS0756 & &07 56 53.11 & +08 58 31.8 & 16.27 & 16.8 &VIII \\   
SDSS0805 & &08 05 34.49 & +07 20 29.1 & 18.52 & 18.5 &VI \\  
SDSS0809 & &08 09 08.39 & +38 14 06.2 & 15.61 & 15.7 &II \\  
SDSS0812 & &08 12 56.85 & +19 11 57.8 & 15.80 & 16.0 &V  \\  
SDSS0813 & &08 13 52.02 & +28 13 17.3 & 17.16 & 17.4 &IV \\  
SDSS0838 & &08 38 45.26 & +49 10 55.1 & 19.59 & 18.5 &I \\  
SDSS0844 & V495 Hya &08 44 00.10 & +02 39 19.3 & 18.34 & 18.7 &II \\
SDSS0912 & &09 12 42.18 & +62 09 40.1 & 18.81 & 19.4 &VIII \\ 
SDSS0916 & HH Cnc &09 16 50.78& +28 49 43.7 & 19.17 & 19.1 &VI \\
SDSS0919 & LV Cnc & 09 19 45.11 & +08 57 10.0 & 18.20 & 18.8 &IV \\
SDSS0922 & & 09 22 29.26 & +33 07 43.6 & 18.44 & 19.3 &IV \\ 
SDSS0945 & & 09 45 58.24 & +29 22 53.2 & 19.10 & 19.1 &VI \\ 
SDSS1013 & & 10 13 23.64 & +45 58 58.9 & 18.86 & 19.3 &III \\
SDSS1019 & AC LMi\tablenotemark{a} & 10 19 47.26 & +33 57 53.6 & 18.39 & 17.8 &VI \\
SDSS1023 & NSV 04838 & 10 23 20.28 & +44 05 09.9 & 18.83 & 18.8  &IV \\
SDSS1028 & Leo5 & 10 28 00.09 & +21 48 13.5 & 16.34 & 16.2 &VII \\ 
SDSS1131 & MR UMa & 11 31 22.39 & +43 22 38.6 & 16.17 & 17.9 &V \\ 
SDSS1132 & & 11 32 15.50 & +62 49 00.4 & 18.49 & 18.4 &III \\ 
SDSS1244 & & 12 44 17.89 & +30 04 01.0 & 18.61 & 19.1 &VII \\ 
SDSS1429 & & 14 29 55.86 & +41 45 16.8 & 17.70 & 18.3 & IV \\
SDSS1504 & Boo1\tablenotemark{b} & 15 04 41.76 & +08 47 52.6 & 19.44 & 19.2 &VII \\ 
SDSS1536 & & 15 36 34.42 & +33 28 51.9 & 19.20 & 19.0 &IV \\ 
SDSS1556 & V493 Ser & 15 56 44.23 & $-$00 09 50.3 & 18.05 & 18.7 & I \\
SDSS1610 & V519 Ser & 16 10 07.49 & +03 52 33.2 & 17.36 & 17.8  &VII \\
SDSS1619 & & 16 19 09.10 & +13 51 45.5 & 18.49 & 17.2 & VII \\ 
SDSS1653 & V1227 Her & 16 53 59.06 & +20 10 10.4 & 17.53 & 18.5 &V \\
SDSS1656 & V1229 Her & 16 56 58.13 & +21 21 39.3 & 18.52 & 18.9 &V \\
SDSS1659 & & 16 59 51.69 & +19 27 45.6 & 16.76 & 17.3 &V \\ 
SDSS1730 & & 17 30 08.38 & +62 47 54.7 & 15.92 & 17.8 &I\\ 
\enddata
\tablecomments{Coordinates and $g$ magnitudes
are taken from the 
discovery papers referred to in the last column.
Synthetic $V$ magnitudes (column 4) are from our
average spectra.  References are as follows:
I -- \citet{szkodyi};
II -- \citet{szkodyii}; 
III -- \citet{szkodyiii}; 
IV -- \citet{szkodyiv}; 
V -- \citet{szkodyv}; 
VI -- \citet{szkodyvi}; 
VII -- \citet{szkodyvii}; 
VIII -- \citet{szkodyviii};
Wils -- \citet{wils2010}
}
\tablenotetext{a}{Also known as HS1016+3412 \citep{aung06}.}
\tablenotetext{b}{As listed in \citet{downescat}.}
\label{tab:star_info}
\end{deluxetable}

\clearpage

\begin{deluxetable}{lrrrrr}
\tablecolumns{6}
\tablewidth{0pt}
\tablecaption{Radial Velocities}
\tablehead{
\colhead{Star} &
\colhead{Time\tablenotemark{a}} &
\colhead{$v_{\rm emn}$} &
\colhead{$\sigma$} & 
\colhead{$v_{\rm abs}$} &
\colhead{$\sigma$} \\
\colhead{SDSS J} &
\colhead{} &
\colhead{[km s$^{-1}$]} &
\colhead{[km s$^{-1}$]} &
\colhead{[km s$^{-1}$]} &
\colhead{[km s$^{-1}$]} \\
}
\startdata
075653.11+085831.8 & 	2455944.9631 & 	 -327 & 	  24&	 \nodata &	 \nodata  \\
075653.11+085831.8 & 	2455944.9713 & 	 -253 & 	  20&	 \nodata &	 \nodata  \\
075653.11+085831.8 & 	2455944.9800 & 	   -7 & 	  14&	 \nodata &	 \nodata  \\
075653.11+085831.8 & 	2455944.9886 & 	  131 & 	  11&	 \nodata &	 \nodata  \\
075653.11+085831.8 & 	2455945.0084 & 	  324 & 	  16&	 \nodata &	 \nodata  \\
075653.11+085831.8 & 	2455945.0170 & 	  347 & 	  16&	 \nodata &	 \nodata  \\
075653.11+085831.8 & 	2455945.0257 & 	  517 & 	  21&	 \nodata &	 \nodata  \\
075653.11+085831.8 & 	2455945.7337 & 	  161 & 	  18&	 \nodata &	 \nodata  \\
075653.11+085831.8 & 	2455945.7396 & 	   13 & 	  17&	 \nodata &	 \nodata  \\
\enddata
\tablenotetext{a}{Barycentric Julian Date of mid-exposure.  The time base is UTC.}
\tablecomments{Table \ref{tab:velocities} is published in its entirety in the electronic 
edition of The Astronomical Journal, A portion is shown here for guidance regarding its form and content.}
\label{tab:velocities}
\end{deluxetable}

\clearpage

\begin{deluxetable}{lllrrcc}
\tablecolumns{7}
\footnotesize
\tablewidth{0pt}
\tablecaption{Fits to Radial Velocities}
\tablehead{
\colhead{Data set} & 
\colhead{$T_0$\tablenotemark{a}} & 
\colhead{$P$} &
\colhead{$K$} & 
\colhead{$\gamma$} & 
\colhead{$N$} &
\colhead{$\sigma$\tablenotemark{b}}  \\ 
\colhead{} & 
\colhead{} &
\colhead{(d)} & 
\colhead{(km s$^{-1}$)} &
\colhead{(km s$^{-1}$)} & 
\colhead{} &
\colhead{(km s$^{-1}$)} \\
}
\startdata

SDSS0756 & 55946.9049(8) & 0.13718(11)\tablenotemark{c} & 376(20) & $ 33(12)$ & 43 &  56 \\[1.2ex]
SDSS0805 abs. & 54479.713(2) & 0.2296(6) &  216(12) & $ 32(9)$ & 16 &  27 \\
SDSS0805 emn. & 54479.615(10) & 0.226(2) &  81(18) & $-9(14)$ & 16 &  37 \\ 
combined      &              &  0.2294(6)  \\[1.2ex]
SDSS0809 & 53071.991(3) & 0.133616(2) &  87(14) & $ 55(10)$ & 158 &  44 \\[1.2ex]
SDSS0812 & 53810.6770(20) & 0.1600(2)\tablenotemark{d} &  172(13) & $-0(9)$ & 78 &  42 \\[1.2ex]
SDSS0813 & 53812.5965(19) & 0.1220(4) &  140(13) & $-84(9)$ & 45 &  31 \\[1.2ex]
SDSS0838 & 56302.6769(17) & 0.07034(19) &  49(7) & $ 71(5)$ & 41 &  25 \\[1.2ex]
SDSS0844 & 53744.033(4) & 0.2070(7) &  47(6) & $ 54(4)$ & 89 &  20 \\[1.2ex]
SDSS0912 & 56299.0564(17) & 0.08013(9) &  101(15) & $-18(10)$ & 46 &  34\\[1.2 ex]
SDSS0916 & 56353.697(3) & 0.1845(5) &  31(3) & $ 8(2)$ & 43 &   7 \\[1.2 ex]
SDSS0919 & 53746.7419(13) & 0.05647(12) &  65(9) & $ 73(6)$ & 64 &  29 \\[1.2ex]
SDSS0922 [WHT] & 54186.3557(12) & 0.06578(11) &  51(6) & $ 76(4)$ & 32 &  19 \\
SDSS0922 [MDM] & 56354.6528(17) & 0.06547(14) &  43(7) & $ 20(5)$ & 49 &  24 \\
combined & & 0.06567(9) & & 52 & & \\[1.2ex]
SDSS0945 & 54526.9796(11) & 0.0639(2) & 255(22) & $ 5(17)$ & 22 &  61 \\[1.2ex]
SDSS1013 & 55280.731(3) & 0.08204(13)  &  42(8) & $ 19(6)$ & 80 &  31 \\[1.2ex]
SDSS1019 & 54125.994(2) & 0.07915(14) &  45(9) & $ 19(6)$ & 31 &  20 \\[1.2ex]
SDSS1023 & 52672.773(2) & 0.0679(3) &  67(12) & $-86(9)$ & 42 &  29 \\[1.2ex]
SDSS1028 & 51191.860(4) & 0.14606(17) &  82(12) & $-75(9)$ & 74 &  32 \\[1.2ex]
SDSS1131 & 52671.8142(15) & 0.06325(10) &  37(6) & $-21(4)$ & 53 &  13 \\[1.2ex]
SDSS1132 & 54525.671(2) & 0.0689(2) &  28(5) & $-16(4)$ & 73 &  17  \\[1.2ex]
SDSS1244 & 54898.7621(17) & 0.07744(9) &  80(11) & $-8(8)$ & 59 &  25 \\[1.2ex]
SDSS1429 & 53549.677(3) & 0.0685(2):\tablenotemark{e} & 86.0(234) & $-41(17)$ & 55 &  55 \\[1.2ex]
SDSS1504 & 55281.0322(19) & 0.08089(18) &  35(6) & $ 19(4)$ & 43 &  13 \\[1.2ex]
SDSS1536 & 56451.675(2) & 0.0921(2) &  88(12) & $-107(8)$ & 34 &  34 \\[1.2ex]
SDSS1556 & 52812.7011(15) & 0.08012(8) &  72(8) & $-31(6)$ & 52 &  26  \\[1.2ex]
SDSS1610 & 52328.0203(16) & 0.1322739(2) &  90(6) & $ 17(5)$ & 93 &  21 \\[1.2ex]
SDSS1619 abs. & 56453.839(3) & 0.2867(6) &  129(10) & $ 79(7)$ & 17 &  25 \\ 
SDSS1619 emn. & 56453.689(3) & 0.2863(6) &  160(10) & $ 19(8)$ & 18 &  27 \\
combined & & 0.2865(5) & \\[1.2ex]
SDSS1653 & 55367.8268(11) & 0.06310(9) &  44(5) & $-1(4)$ & 34 &  16 \\[1.2ex]
SDSS1656 & 53903.977(3) & 0.06733(14) &  45(13) & $-15(3)$ & 67 &  42 \\[1.2ex]
SDSS1659 & 54277.655(4) & 0.141(2) &  101(18) & $-15(12)$ & 106 &  54 \\[1.2ex]
SDSS1730 & 52438.943(2) & 0.07657(16) &  56(10) & $-56(7)$ & 39 &  24 \\[1.2ex]
\enddata
\tablecomments{Parameters of least-squares sinusoid fits to the radial
velocities, of the form $v(t) = \gamma + K \sin(2 \pi(t - T_0)/P$.}
\tablenotetext{a}{Heliocentric Julian Date minus 2400000.  The epoch is chosen
to be near the center of the time interval covered by the data, and
within one cycle of an actual observation.}
\tablenotetext{b}{RMS residual of the fit.}
\tablenotetext{c}{The eclipse period, refined from CRTTS data, is 
0.1369739(2) d (see text).}
\tablenotetext{d}{The eclipse period, refined from CRTTS data, is 0.1600525(3) d 
(see text).}
\tablenotetext{e}{The daily cycle count for SDSS1429 is ambiguous; see text
for details.
.}
\label{tab:parameters}
\end{deluxetable}

\addtocounter{figure}{-1}
\begin{figure}
\caption{Average spectra, periodograms, and folded velocity
curves for SDSS0756, SDSS0805, SDSS0812, and SDSS0809.
The vertical scales, unlabeled to save space, are (1)
for the spectra, $f_\lambda$ in units of $10^{-16}$ erg
s$^{-1}$ cm$^{-2}$ \AA$^{-1}$; (2) for the periodograms,
$1 / \chi^2$ (dimensionless); and (3) for the radial velocity
curves, barycentric radial velocity in km s$^{-1}$.  
For SDSS0805, the lower trace shows the remainder after
subtracting the scaled spectrum of a K4 star (see text).  
In cases where velocities are from more than one observing run, 
the periodogram is labeled with the word ``peaks'', because the 
curve shown formed by joining local maxima in the 
full periodogram with straight lines. This suppresses fine-scale
ringing due to the unknown number of cycle counts between runs. 
The folded velocity curves all show the same data plotted over
two cycles for continuity, and the best-fit sinusoid (see 
\ref{tab:parameters}) is also plotted.  The velocities shown
are plot H$\alpha$ emission velocities, but in SDSS0805 cross-correlation
velocities of the secondary star are also plotted; they are shown
in red in the online version, and are distinguished by their
smaller error bars and larger velocity amplitude.
}
\end{figure}

\begin{figure}
\plotone{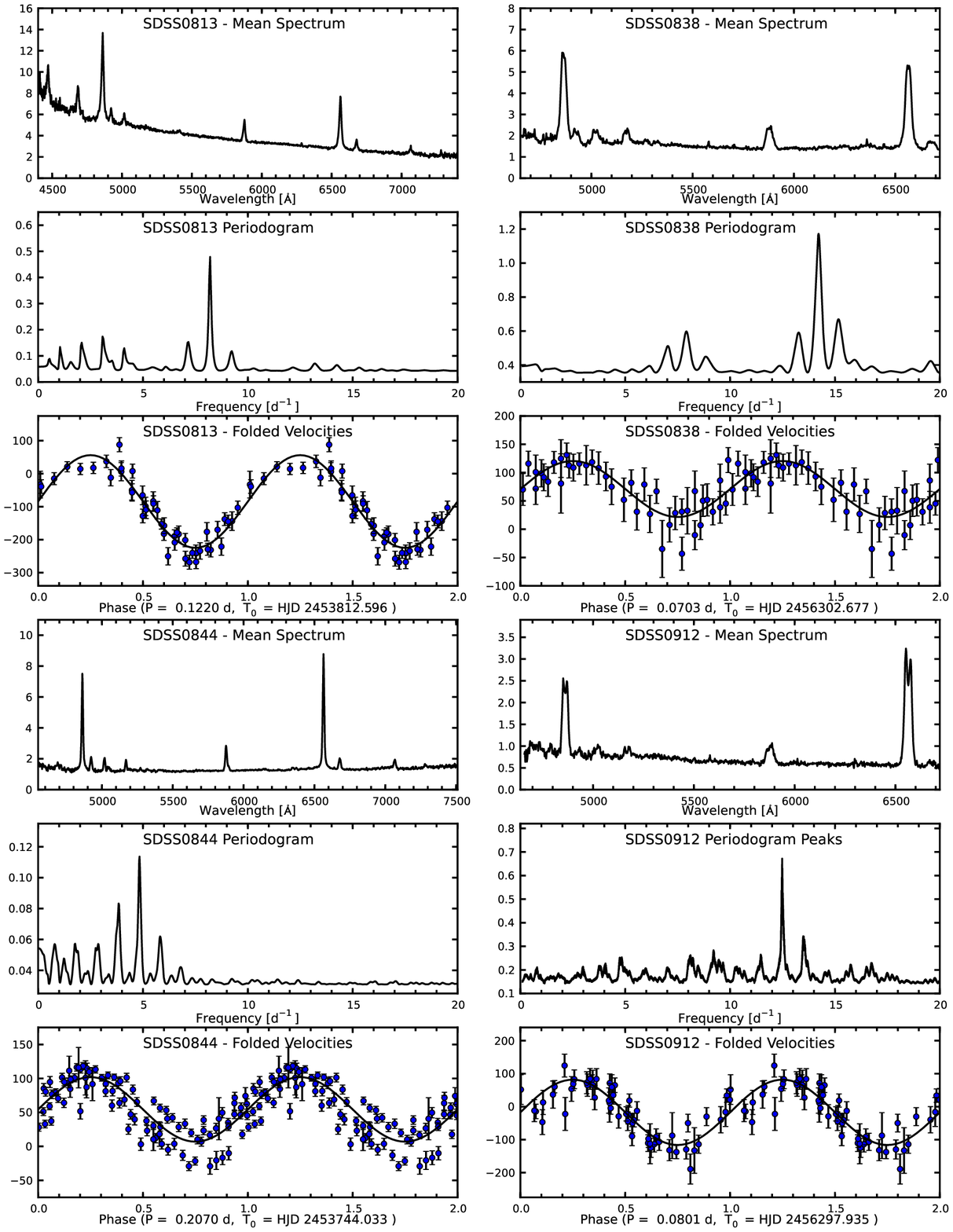}
\vspace{0.4 truein}
\caption{Similar to Fig.~\ref{fig:cvplot1}, but for 
SDSS0813, SDSS0838, SDSS0844, and SDSS0912.
}
\label{fig:cvplot2}
\end{figure}

\begin{figure}
\plotone{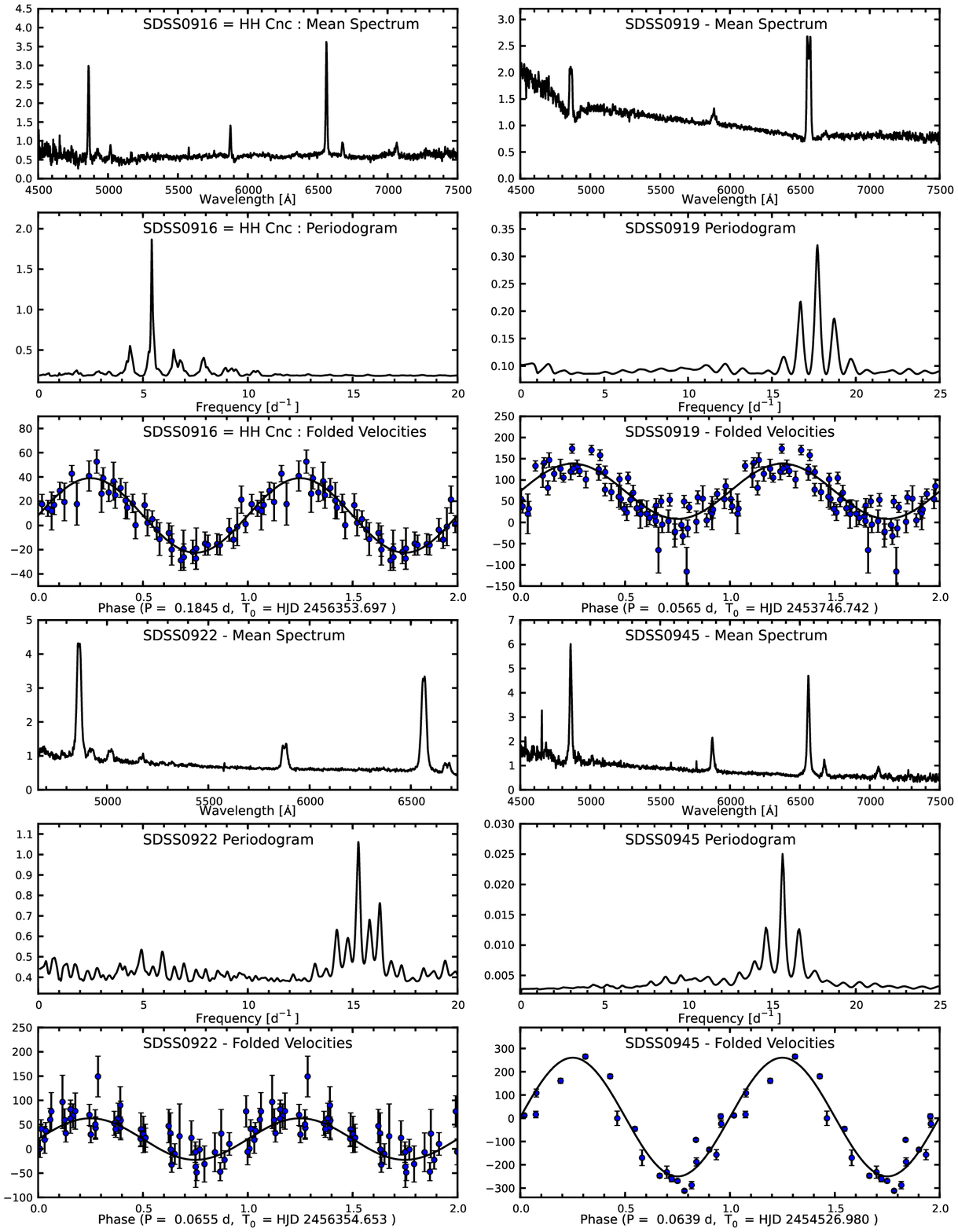}
\vspace{0.4 truein}
\caption{Similar to Fig.~\ref{fig:cvplot1}, but for 
SDSS0916 = HH Cnc, SDSS0919, SDSS0922, and SDSS0945.
}
\label{fig:cvplot3}
\end{figure}

\begin{figure}
\plotone{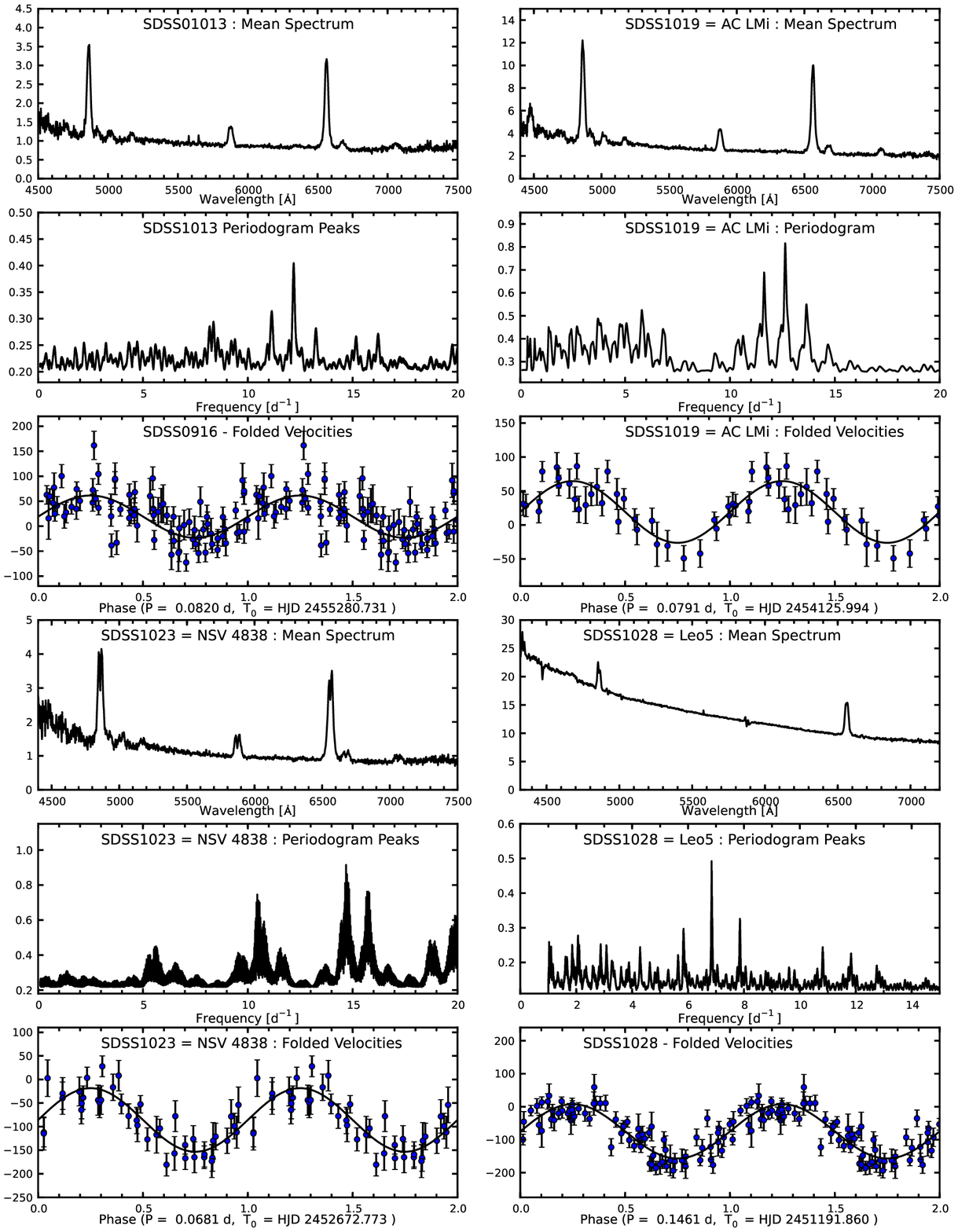}
\vspace{0.4 truein}
\caption{Similar to Fig.~\ref{fig:cvplot1}, but for 
SDSS1013, SDSS1019 = AC LMi, SDSS1023 = NSV 4838, and SDSS1028 = Leo5.
}
\label{fig:cvplot4}
\end{figure}

\begin{figure}
\plotone{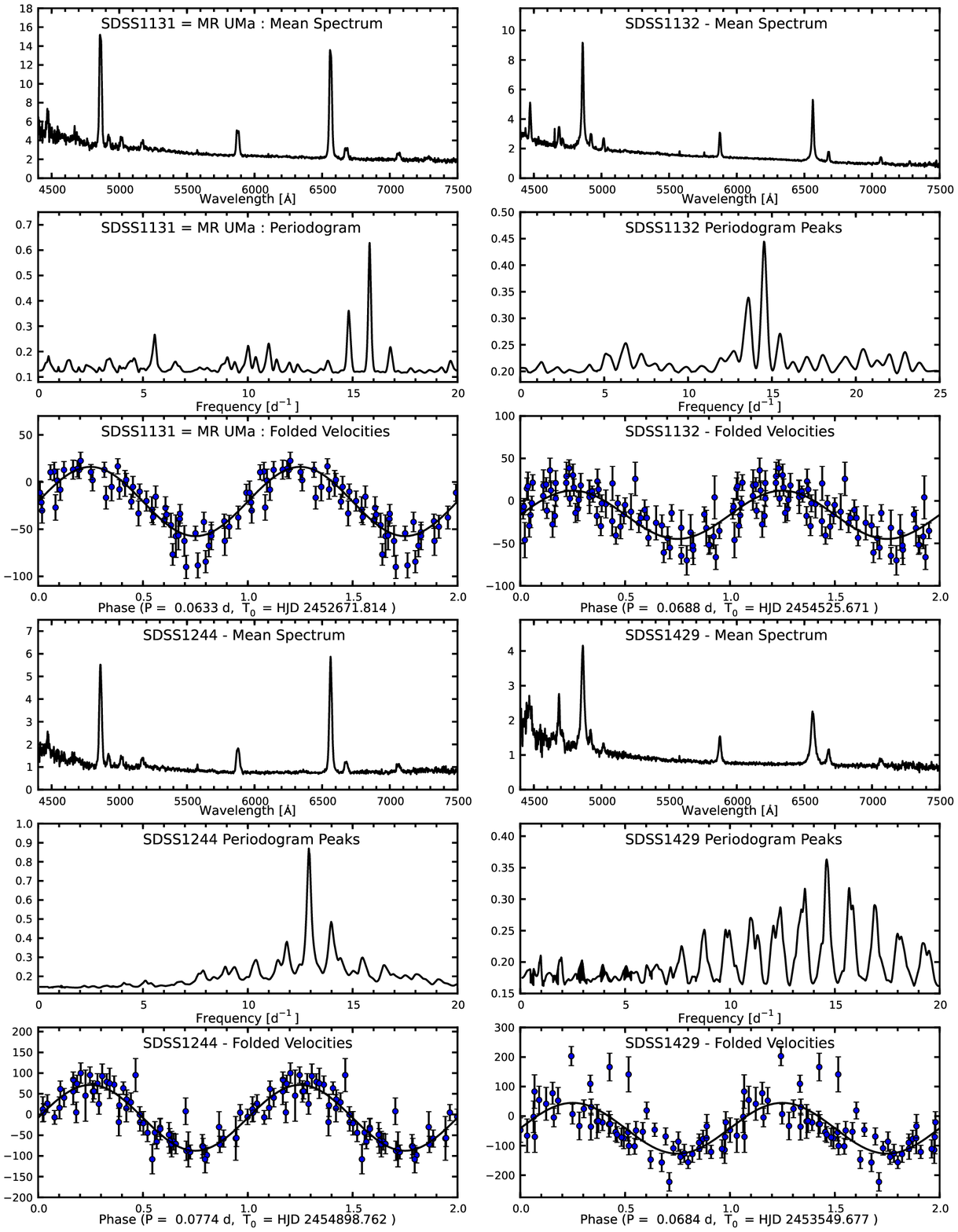}
\vspace{0.4 truein}
\caption{Similar to Fig.~\ref{fig:cvplot1}, but for 
SDSS1131 = MR UMa, SDSS1132, SDSS1244, and SDSS1429.
}
\label{fig:cvplot5}
\end{figure}

\begin{figure}
\plotone{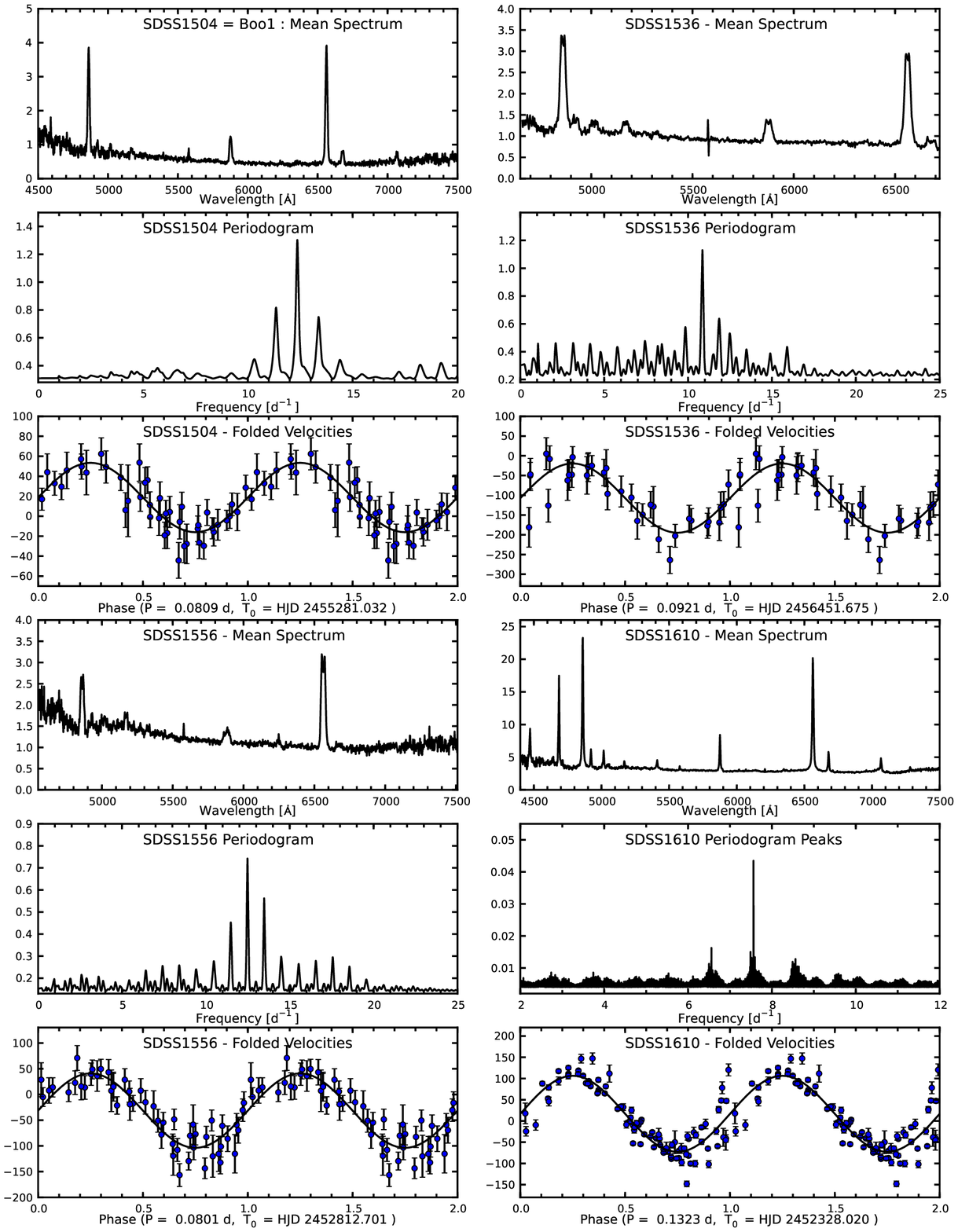}
\vspace{0.4 truein}
\caption{Similar to Fig.~\ref{fig:cvplot1}, but for 
SDSS1504 = Boo1, SDSS1536, SDSS1556, and SDSS1610.
}
\label{fig:cvplot6}
\end{figure}

\begin{figure}
\plotone{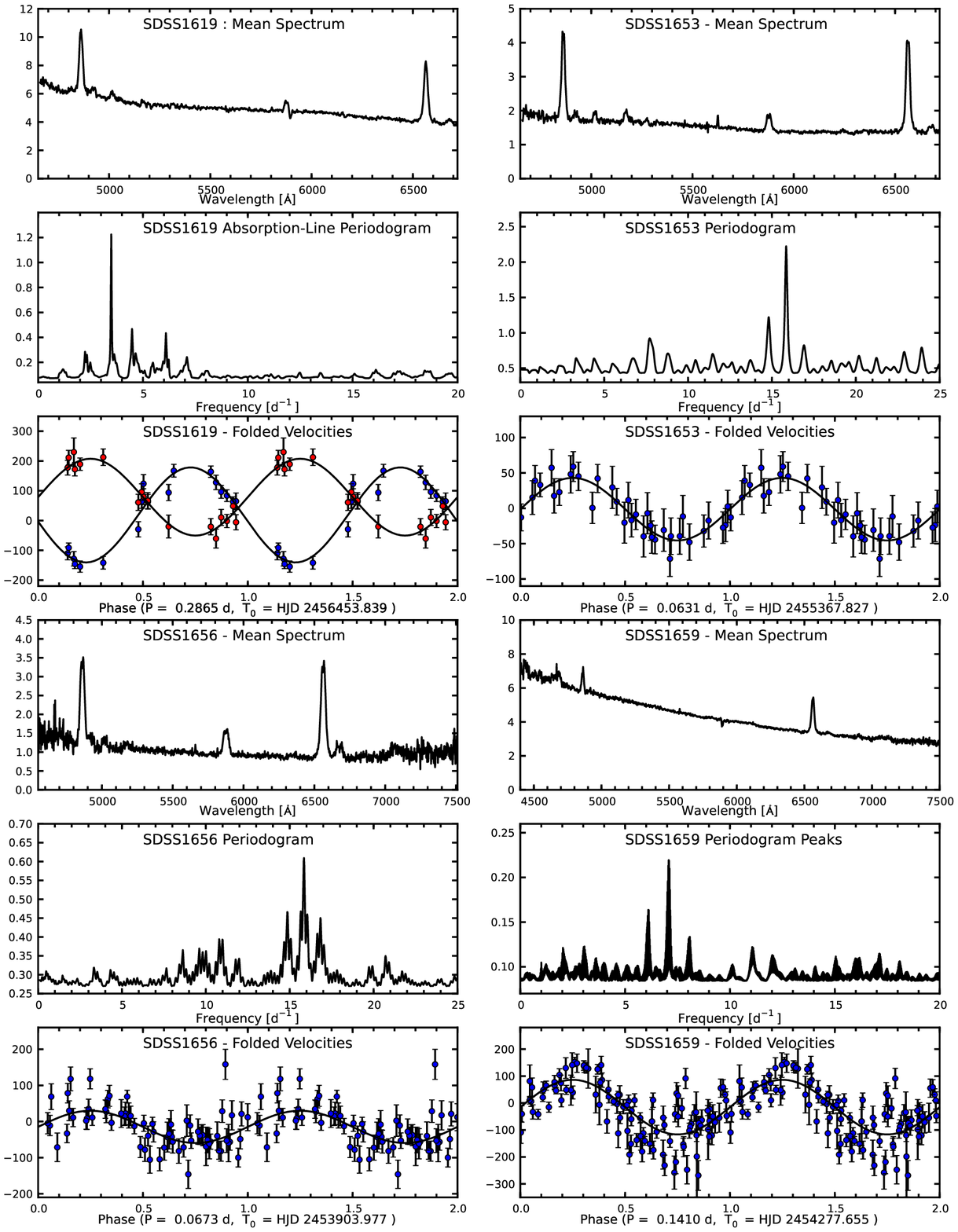}
\vspace{0.4 truein}
\caption{Similar to Fig.~\ref{fig:cvplot1}, but for 
SDSS1619, SDSS1653, SDSS1656, and SDSS1659.  For SDSS1619,
both emission and absorption velocities 
are plotted. 
}
\label{fig:cvplot7}
\end{figure}

\begin{figure}
\plotone{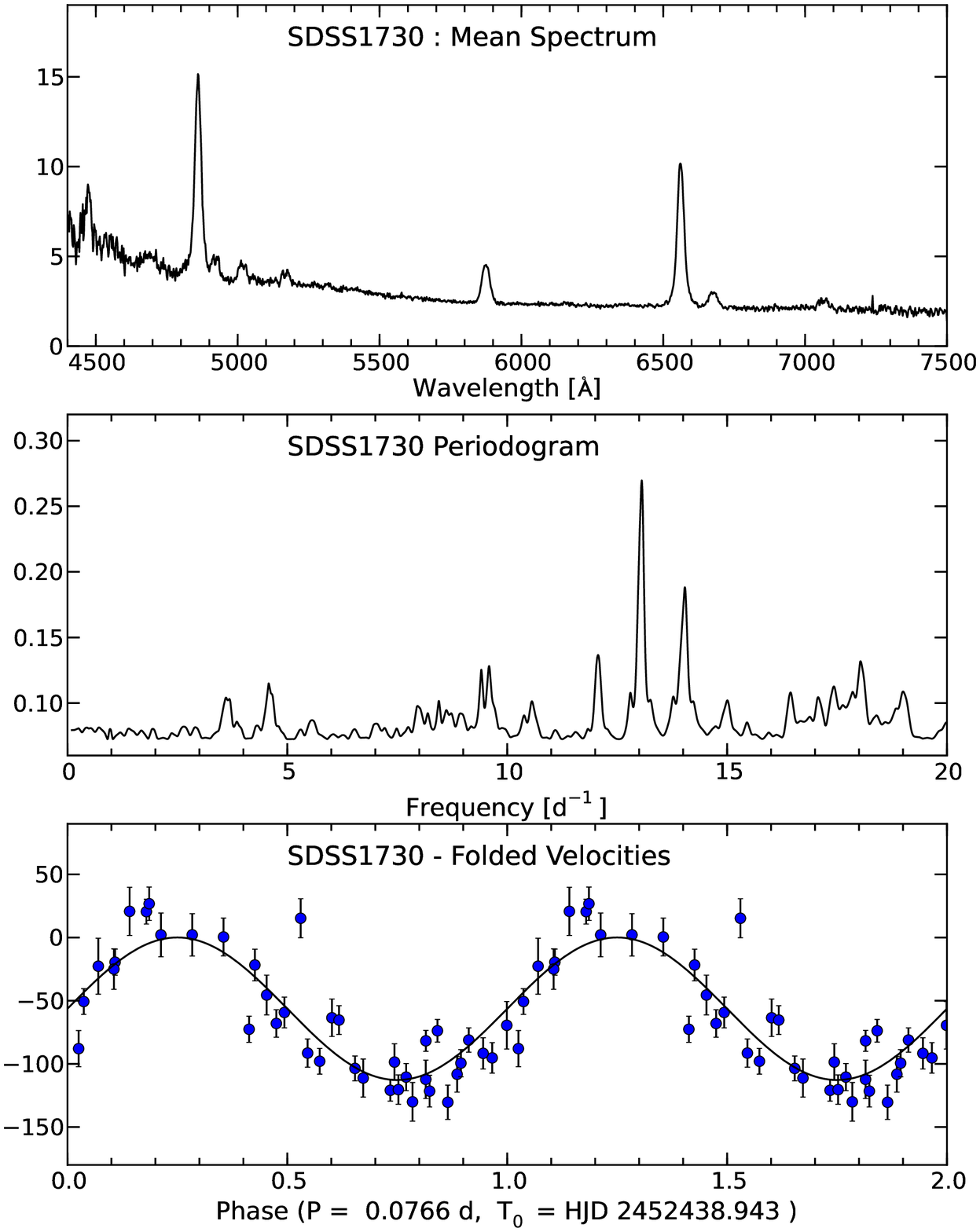}
\caption{Similar to Fig.~\ref{fig:cvplot1}, but for 
SDSS1730.
}
\label{fig:cvplot8}
\end{figure}

\begin{figure}
\epsscale{0.9}
\plotone{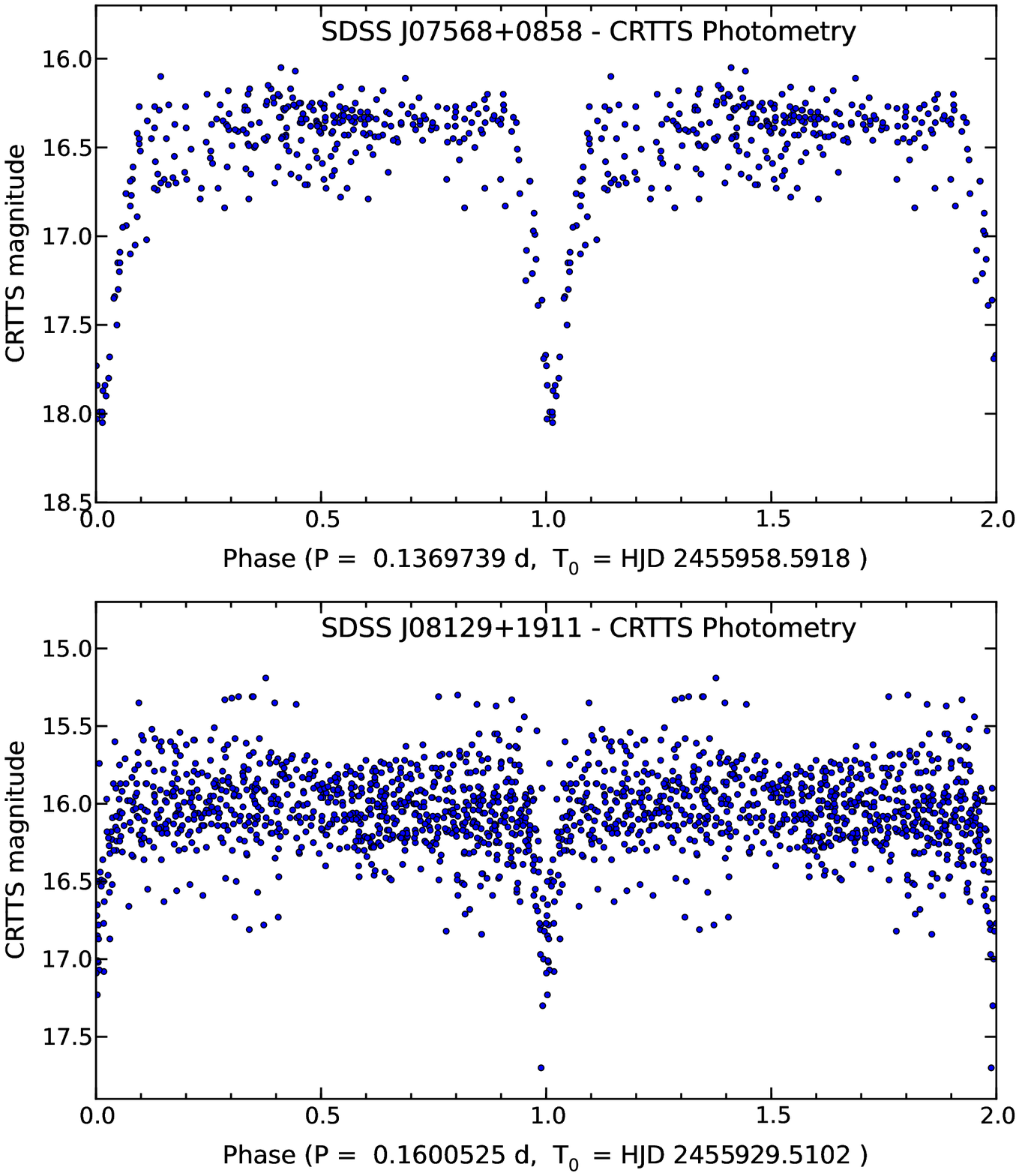}
\caption{Archival magnitudes of SDSS0756 and SDSS0812 from CRTTS,
folded using refined values of the eclipse period.
}
\label{fig:crtscurves}
\end{figure}

\begin{figure}
\plotone{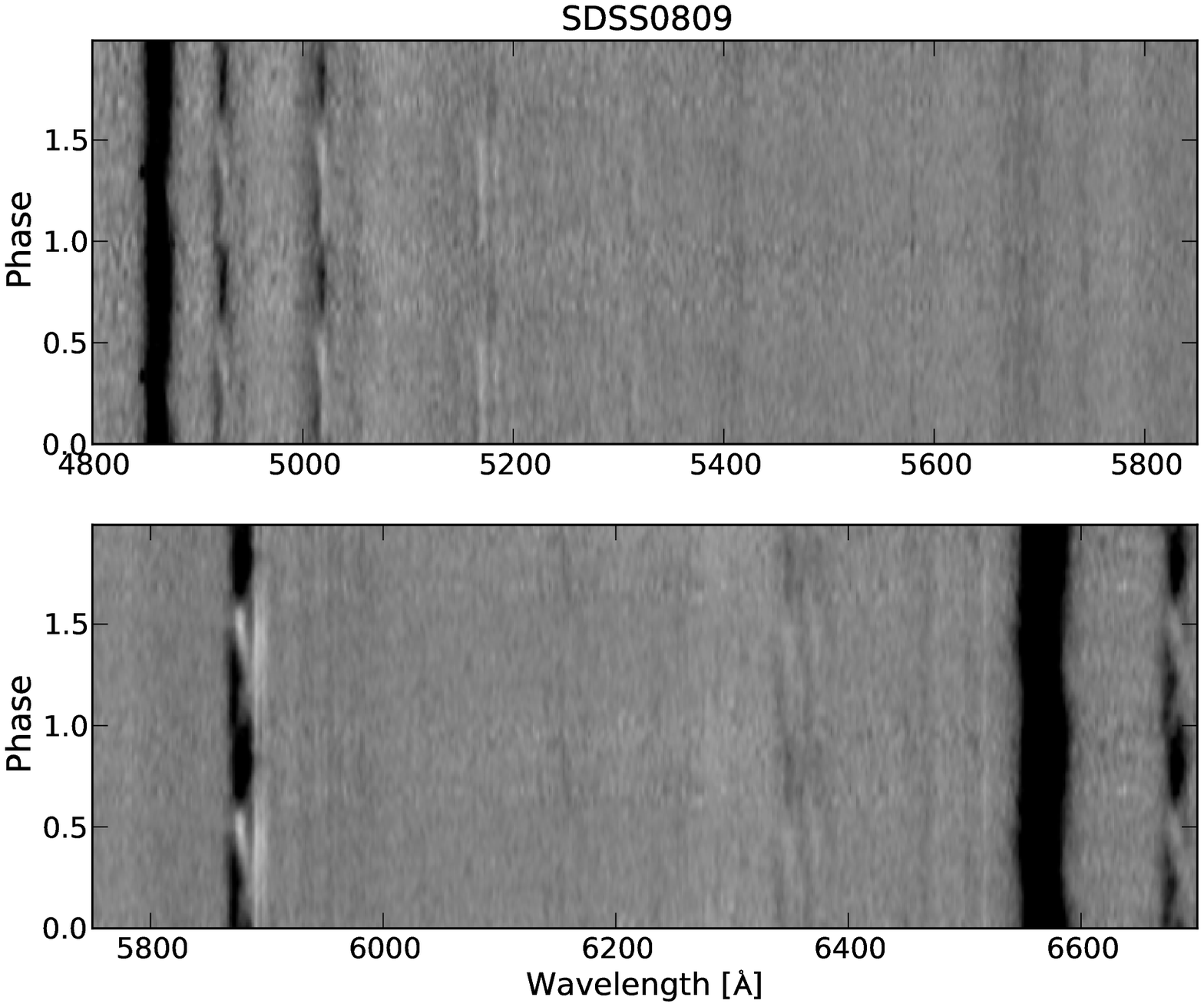}
\caption{A greyscale representation of the rectified spectra of
SDSS0809.  Orbital phase increases upward; each line in the image 
is the average of spectra near that orbital phase, smoothed 
with a Gaussian over phase.  The greyscale is negative (black = bright),
from 0.8 to 1.3 times the continuum in order to bring out faint 
features.  The zero of phase is from the eclipse ephemeris from
\citet{rodrig07}.}
\label{fig:sdss0809trail}
\end{figure}


\clearpage

\begin{thebibliography}

\bibitem[Atlee \& Gould(2007)]{atlee07} Atlee, D.~W., \& Gould, A.\ 2007, \apj, 664, 53 

\bibitem[Aungwerojwit et al.(2006)]{aung06} Aungwerojwit, A., G{\"a}nsicke, B.~T., Rodr{\'{\i}}guez-Gil, P., et al.\ 2006, \aap, 455, 659 

\bibitem[Augusteijn et al.(1996)]{v485cen} Augusteijn, T., 
van der Hooft, F., de Jong, J.~A., \& van Paradijs, J.\ 1996, \aap, 311, 889 

\bibitem[Baraffe \& Kolb(2000)]{bk00} Baraffe, I., Kolb, U. 2000,
\mnras, 318, 354

\bibitem[Barker \& Kolb(2003)]{barker03} Barker, J., 
\& Kolb, U.\ 2003, \mnras, 340, 623  

\bibitem[Bessell(1990)]{bessell} Bessell, M. S. 1990, \pasp, 102, 1181

\bibitem[Beuermann et al.(1999)]{beuermann99} Beuermann, K., 
Baraffe, I., \& Hauschildt, P.\ 1999, \aap, 348, 524 

\bibitem[Bochanski et al.(2007)]{bochanskitemplate} Bochanski, J.~J., 
West, A.~A., Hawley, S.~L., \& Covey, K.~R.\ 2007, \aj, 133, 531 

\bibitem[Boeshaar(1976)]{boeshaar} Boeshaar, P. 1976, Ph.~D. Thesis, 
The Ohio State University

\bibitem[Breedt et al.(2014)]{breedt14} Breedt, E., 
G{\"a}nsicke, B.~T., Drake, A.~J., et al.\ 2014, \mnras, 443, 3174 


\bibitem[Dillon et al.(2008)]{dillon08} Dillon, M., et al.\ 
2008, \mnras, 386, 1568 

\bibitem[Downes et al.(2001)]{downescat} Downes, R.~A., Webbink, 
R.~F., Shara, M.~M., et al.\ 2001, \pasp, 113, 764 

\bibitem[Drake et al.(2009)]{drakecrtts} Drake, A.~J., et al.\ 
2009, \apj, 696, 870 

\bibitem[Dworetsky(1983)]{dworetsky83} Dworetsky, M.~M.\ 1983, 
\mnras, 203, 917 

\bibitem[Filippenko(1982)]{filippenko} Filippenko, A.~V.\ 1982, 
\pasp, 94, 715 

\bibitem[Filippenko et al.(1985)]{fili85} Filippenko, A.~V., 
Sargent, W.~L.~W., \& Hazard, C.\ 1985, \pasp, 97, 41 

\bibitem[G{\"a}nsicke et al.(2009)]{unveils} G{\"a}nsicke, 
B.~T., et al.\ 2009, \mnras, 397, 2170 

\bibitem[G{\"u}lsecen \& Eseno{\~g}lu(2014)]{gulsecen13} 
G{\"u}lsecen, H., \& Eseno{\~g}lu, H.\ 2014, \na, 28, 49 

\bibitem[Horne(1986)]{horne} Horne, K.\ 1986, \pasp, 98, 609 

\bibitem[Honeycutt \& Kafka(2004)]{honeycuttkafka} 
Honeycutt, R.~K., \& Kafka, S.\ 2004, \aj, 128, 1279 

\bibitem[Imada et al.(2009)]{imada09} Imada, A., Yasuda, T., 
Omodaka, T., et al.\ 2009, \pasj, 61, 535  

\bibitem[Jiang et al.(2000)]{jiang2000} Jiang, X.~J., Engels, 
D., Wei, J.~Y., Tesch, F., \& Hu, J.~Y.\ 2000, \aap, 362, 263  

\bibitem[Kato et al.(2009)]{kato09} Kato, T., et al.\ 2009, 
\pasj, 61, 395 

\bibitem[Kato et al.(2010)]{kato10} Kato, T., Maehara, H., 
Uemura, M., et al.\ 2010, \pasj, 62, 1525 


\bibitem[Keenan \& McNeil(1989)]{keenan} Keenan, P.~C., \& McNeil, 
R.~C.\ 1989, \apjs, 71, 245 

\bibitem[Knigge(2006)]{kniggedonor} Knigge, C.\ 2006, \mnras, 373,
484

\bibitem[Kurtz \& Mink(1998)]{kurtzmink} Kurtz, M.~J. \& Mink,
D.~J.\ 1998, \pasp, 110, 934

\bibitem[Linnell et al.(2007)]{linnell07} Linnell, A.~P., Hoard, 
D.~W., Szkody, P., et al.\ 2007, \apj, 654, 1036 

\bibitem[Lipunov et al.(2010)]{lipunovmaster} Lipunov, V., Kornilov, 
V., Gorbovskoy, E., et al.\ 2010, Advances in Astronomy, 2010,  

\bibitem[Littlefair et al.(2007)]{sdss1507litt} Littlefair, S.~P., 
Dhillon, V.~S., Marsh, T.~R., G{\"a}nsicke, B.~T., Baraffe, I., 
\& Watson, C.~A.\ 2007, \mnras, 381, 827 

\bibitem[Monet et al.(2003)]{usnob} Monet, D.~G., Levine, 
S.~E., Canzian, B., et al.\ 2003, \aj, 125, 984 

\bibitem[Mukadam et al.(2007)]{mukadam07} Mukadam, A.~S., 
G{\"a}nsicke, B.~T., Szkody, P., Aungwerojwit, A., Howell, S.~B., Fraser, 
O.~J., \& Silvestri, N.~M.\ 2007, \apj, 667, 433 

\bibitem[Munari et al.(1997)]{munari97} Munari, U., Zwitter, T., 
\& Bragaglia, A.\ 1997, \aaps, 122, 495   


\bibitem[Patterson(1984)]{patterson84} Patterson, J.\ 1984, \apjs, 
54, 443 

\bibitem[Patterson(1998)]{pattlate} Patterson, J.\ 1998, \pasp, 
110, 1132 

\bibitem[Patterson et al.(2002)]{pattwzsge} Patterson, J., Masi, 
G., Richmond, M.~W., et al.\ 2002, \pasp, 114, 721 


\bibitem[Patterson et al.(2008)]{sdss1507patt} Patterson, J., 
Thorstensen, J.~R., \& Knigge, C.\ 2008, \pasp, 120, 510 


\bibitem[Rodrigues et al.(2006)]{rodrigues06} Rodrigues, C.~V., 
Jablonski, F.~J., D'Amico, F., et al.\ 2006, \mnras, 369, 1972   

\bibitem[Rodr{\'{\i}}guez-Gil et al.(2007)]{rodrig07} 
Rodr{\'{\i}}guez-Gil, P., G{\"a}nsicke, B.~T., Hagen, H.-J., et al.\ 2007, 
\mnras, 377, 1747 

\bibitem[Samus et al.(2012)]{gcvs} Samus, N.~N., Durlevich, 
O.~V., \& et al.\ 2012, General Catalogue of Variable Stars Database,
Institute of Astronomy of Russian Academy of Sciences and Sternberg
State Astronomical Institute of the Moscow State University 

\bibitem[Schlegel, Finkbeiner, \& Davis(1998)]{schlegel98}
Schlegel, D. J., Finkbeiner, D. P., \& Davis, M. 1998, \apj, 500, 525

\bibitem[Schneider \& Young(1980)]{sy80} Schneider, D. and Young, P. 1980,
\apj, 238, 946

\bibitem[Schwope et al.(1999)]{schwope99} Schwope, A.~D.,
Schwarz, R., Staude, A., et al.\ 1999, Annapolis Workshop on Magnetic
Cataclysmic Variables (ed. C. Hellier \& K. Mukai), ASPC, 157, 71

\bibitem[Shappee et al.(2014)]{shappee14} Shappee, B.~J., Prieto, 
J.~L., Grupe, D., et al.\ 2014, \apj, 788, 48 

\bibitem[Southworth et al.(2007)]{southworth07} Southworth, J., 
Marsh, T.~R., G{\"a}nsicke, B.~T., Aungwerojwit, A., Hakala, P., de 
Martino, D., \& Lehto, H.\ 2007, \mnras, 382, 1145 

\bibitem[Southworth et al.(2008)]{southworth08} Southworth, J., et 
al.\ 2008, \mnras, 391, 591 

\bibitem[Szkody et al.(2000)]{szkodygw} Szkody, P., Desai, V., 
\& Hoard, D.~W.\ 2000, \aj, 119, 365 

\bibitem[Szkody et al.(2002)]{szkodyi} Szkody, P., et al.\ 
2002, \aj, 123, 430 

\bibitem[Szkody et al.(2003)]{szkodyii} Szkody, P., et al.\ 
2003, \aj, 126, 1499 

\bibitem[Szkody et al.(2004)]{szkodyiii} Szkody, P., et al.\ 
2004, \aj, 128, 1882 

\bibitem[Szkody et al.(2005)]{szkodyiv} Szkody, P., et al.\ 
2005, \aj, 129, 2386 

\bibitem[Szkody et al.(2006)]{szkodyv} Szkody, P., et al.\ 
2006, \aj, 131, 973 

\bibitem[Szkody et al.(2007)]{szkodyvi} Szkody, P., et al.\ 
2007, \aj, 134, 185 

\bibitem[Szkody et al.(2009)]{szkodyvii} Szkody, P., et al.\ 
2009, \aj, 137, 4011 

\bibitem[Szkody et al.(2011)]{szkodyviii} Szkody, P., Anderson, 
S.~F., Brooks, K., et al.\ 2011, \aj, 142, 181 


\bibitem[Taylor(1999)]{taylorphd} Taylor, C. J. 1999, PhD Thesis, Dartmouth College

\bibitem[Thorstensen \& Freed(1985)]{tf85} Thorstensen, J.~R., \& Freed, I.~W.\ 1985, \aj, 90, 2082 

\bibitem[Thorstensen et al.(1991)]{thorstensen91} Thorstensen, J.~R., 
Ringwald, F.~A., Wade, R.~A., Schmidt, G.~D., 
\& Norsworthy, J.~E.\ 1991, \aj, 102, 272 

\bibitem[Thorstensen et al.(2002)]{eipsc} Thorstensen, J.~R., 
Fenton, W.~H., Patterson, J.~O., Kemp, J., Krajci, T., 
\& Baraffe, I.\ 2002, \apjl, 567, L49 

\bibitem[Thorstensen et al.(1996)]{tpst} Thorstensen, J. R., Patterson, J.,
Thomas, G., \& Shambrook, A. 1996, \pasp, 108, 73

\bibitem[Thorstensen et al.(2002)]{thorshortest} Thorstensen, J.~R., 
Patterson, J., Kemp, J., \& Vennes, S.\ 2002, \pasp, 114, 1108 

\bibitem[Thorstensen(2003)]{thorparallax1} Thorstensen, J.~R.\ 2003, 
\aj, 126, 3017 

\bibitem[Thorstensen et al.(2008)]{thorparallax2} Thorstensen, J.~R., 
L{\'e}pine, S., \& Shara, M.\ 2008, \aj, 136, 2107 

\bibitem[Thorstensen \& Skinner(2012)]{thorcrts} Thorstensen, J.~R., \& Skinner, J.~N.\ 2012, \aj, 144, 81 

\bibitem[Tonry \& Davis(1979)]{tonrydavis79} Tonry, J., \& Davis, 
M.\ 1979, \aj, 84, 1511 

\bibitem[Tovmassian et al.(2014)]{tov2014} Tovmassian, G., 
Hernandez, M. S., Gonz\'alez-Buitrago, D., Zharikov, S., \&
Garc\'ia-D\'iaz, M. T., 2014, \aj, 147, 68

\bibitem[Voges et al.(2000)]{vogesrosatfaint} Voges, W., et al.\ 2000, 
\iaucirc, 7432, 1 

\bibitem[Warner(1995)]{warner95} Warner, B., in {\it Cataclysmic Variable
Stars}, 1995, Cambridge University Press, New York

\bibitem[Wei et al.(1997)]{wei97} Wei, J.-y., Cao, L., Xu, D.-w., 
Hu, J.-y., \& Li, Q.-b.\ 1997, \caa, 21, 146   

\bibitem[Wils et al.(2010)]{wils2010} Wils, P., G{\"a}nsicke, 
B.~T., Drake, A.~J., \& Southworth, J.\ 2010, \mnras, 402, 436  

\bibitem[Wils et al.(2011)]{wils11} Wils, P., Krajci, T., 
\& Simonsen, M.\ 2011, Journal of the American Association of 
Variable Star Observers (JAAVSO), 39, 77 

\bibitem[Woudt et al.(2004)]{woudt04} Woudt, P.~A., Warner, B., 
\& Pretorius, M.~L.\ 2004, \mnras, 351, 1015 

\bibitem[Woudt et al.(2012)]{woudt12} Woudt, P.~A., Warner, B., 
de Bud{\'e}, D., et al.\ 2012, \mnras, 421, 2414 

\bibitem[van Zyl et al.(2004)]{vanzyl04} van Zyl, L., et al.\ 
2004, \mnras, 350, 307 

\end{thebibliography}
\end{document}